\documentclass[12pt,letterpaper]{article}
\pdfoutput=1

\usepackage{color}
\usepackage{slashed}
\usepackage{float}
\definecolor{refkey}{rgb}{1,0,0}
\definecolor{labelkey}{rgb}{0,0,1}
\usepackage{bbold,amsfonts,amssymb,amsmath}
\usepackage{esint}
\usepackage{rotating}
\usepackage[utf8]{inputenc}

\usepackage{graphicx}
\usepackage{caption}
\usepackage{subcaption}
\usepackage{booktabs}

\usepackage{listings}

\usepackage{hyperref}

\newcommand{\be}{\begin{equation}}
\newcommand{\ee}{\end{equation}}
\newcommand{\ben}{\begin{displaymath}}
\newcommand{\een}{\end{displaymath}}
\newcommand{\bea}{\begin{equation}\begin{aligned}}
\newcommand{\eea}{\end{aligned}\end{equation}}
\newcommand{\bean}{\begin{eqnarray*}}
\newcommand{\eean}{\end{eqnarray*}}

\def\a {\alpha}

\newcommand{\bra}[1]{\mbox{$\langle #1 |$}}
\newcommand{\ket}[1]{\mbox{$| #1 \rangle$}}

\newcommand{\eg}{{\it e.g.}}
\newcommand{\ie}{{\it i.e.}}

\newcommand{\tr}{\mbox{Tr}}

\newcommand{\commentout}[1]{}

\newcommand{\beq}{\begin{equation}}
\newcommand{\eeq}{\end{equation}}
\newcommand{\beqr}{\begin{displaymath}}
\newcommand{\eeqr}{\end{displaymath}}
\newcommand{\beqa}{\begin{eqnarray}}
\newcommand{\eeqa}{\end{eqnarray}}
\newcommand{\beqar}{\begin{eqnarray*}}
\newcommand{\eeqar}{\end{eqnarray*}}
\newcommand{\cM}{{\cal M}}

\newcommand{\cO}{{\cal O}}
\newcommand{\cF}{{\cal F}}
\newcommand{\cL}{{\cal L}}

\newcommand{\non}{\nonumber}

\newcommand{\half}{\ensuremath{\frac{1}{2}}}

\newcommand{\cotan}{\ensuremath{\mbox{cotan}}}

\renewcommand{\Im}{\ensuremath{\mathrm{Im}}}

\newcommand{\intxy}{\ensuremath{\int_4^\infty\!\!\!\!\!\! dx \int_4^\infty\!\!\!\!\!\! dy\,}}
\newcommand{\intx}{\ensuremath{\int_4^\infty\!\!\!\!\!\! dx\,}}

\newcommand{\intP}{\ensuremath{\int_{-1}^{+1}\!\!\!\!\!\! d\mu\, P_\ell(\mu)\,}}

\newcommand{\lqcd}{\ensuremath{\Lambda_{\mbox{\scriptsize QCD}}}}
\newcommand{\MeV}{\ensuremath{\mbox{MeV}}}
\newcommand{\GeV}{\ensuremath{\mbox{GeV}}}

\begin{document}

\title{\Large \bf Bootstrapping gauge theories}

\author{
	Yifei He$^\text{1}$,
	Martin Kruczenski$^\text{2}$ \thanks{E-mail: \texttt{yifei.he@ens.fr, markru@purdue.edu.}} \\
[2.0mm]
$^1$ \small Laboratoire de Physique de l'\'Ecole Normale Supérieure, ENS, Université PSL,\\
\small CNRS, Sorbonne Université, Université Paris Cité, F-75005 Paris, France \\
$^2$ \small Department of Physics and Astronomy and PQSEI\thanks{Purdue Quantum Science and Engineering Institute}, \\
\small Purdue University, West Lafayette, IN 47907, USA.}

\date{\today}

\maketitle

\begin{abstract}
 We consider asymptotically free gauge theories with gauge group $SU(N_c)$ and $N_f$ quarks with mass $m_q\ll \lqcd$ that undergo chiral symmetry breaking and confinement. We propose a bootstrap method to compute the S-matrix of the pseudo-Goldstone bosons (pions) that dominate the low energy physics. For the important case of $N_c=3$, $N_f=2$, a numerical implementation of the method gives the phase shifts of the $S0$, $P1$ and $S2$ waves in good agreement with experimental results. The method incorporates gauge theory information ($N_c$, $N_f$, $m_q$, $\lqcd$) by using the form-factor bootstrap recently proposed by Karateev, Kuhn and Penedones together with a finite energy version of the SVZ sum rules. At low energy we impose constraints from chiral symmetry breaking. The only low energy numerical inputs are the pion mass $m_\pi$ and the quark and gluon condensates.
\end{abstract}

\clearpage

\tableofcontents

\newpage




\section{Introduction}

 In this paper we consider the well known problem of finding the low energy physics of an asymptotically free gauge theory with light quarks that undergoes confinement and chiral symmetry breaking. In such case the low energy theory is described by a scalar field theory of pions, the pseudo-Goldstone bosons of chiral symmetry breaking. More concretely, the main question is if one can use high energy data to bootstrap the low-energy pion scattering matrix using a few or even no low-energy parameters.  
 
  In fact, even before gauge theories were known to describe the strong interactions, the S-matrix bootstrap \cite{Eden:1966dnq,chew1966analytic} was already proposed and studied in the 60s and 70s as a way to understand the strong interactions by determining the space of pion scattering matrices that satisfied the constraints of unitarity, analyticity and crossing. The hope was that, under certain conditions, the S-matrix was unique and could be determined in a self-consistent way without resorting to an underlying quantum field theory or Hamiltonian. Part of this idea was that the strong interactions were perhaps the strongest possible coupled theory. However it was later understood that, as pseudo Goldstone bosons, pions are derivatively coupled and as such are weakly coupled at low energies. Later, when the strong interactions were found to be described by an asymptotically free theory, the bootstrap program was largely abandoned except for the case of 2d integrable theories (for example see \cite{Zamolodchikov:1978xm}). Recently, inspired by the success of the conformal bootstrap \cite{Rattazzi:2008pe}, the S-matrix bootstrap proposal was revived by Paulos, Penedones, Toledo, van Rees and Vieira in \cite{Paulos:2016fap,Paulos:2016but,Paulos:2017fhb} leading to a renewed activity in this field as summarized for example in \cite{Kruczenski:2022lot}. In that initial series of papers \cite{Paulos:2016fap,Paulos:2016but,Paulos:2017fhb} it was determined that by maximizing linear functionals in the space of allowed S-matrices, interesting theories can be found and the S-matrix computed numerically. In 2d the sine-Gordon model appears in that way \cite{Paulos:2016fap,Paulos:2016but}. Later, in \cite{He:2018uxa,Cordova:2018uop,Paulos:2018fym} 2d models with global $O(N)$ symmetry were studied and, in particular, it was argued in \cite{He:2018uxa} that the infinite dimensional space of all allowed two-to-two S-matrices is convex with a vertex where the integrable $O(N)$ nonlinear sigma model sits. Later, in \cite{Cordova:2019lot} two- and three-dimensional projections of that space were plotted and the vertex was clearly seen as a kink in the boundary curve defining the allowed space. In 3+1 dimensions \cite{Paulos:2017fhb}, the approach also lead to well determined S-matrices that extremized the coupling but the underlying quantum field theory was not identified. In that way, it becomes clear that in 3+1 dimensions, without introducing extra input, it is not possible to find a particular field theory within the space of S-matrices. In that respect, pion scattering was studied from a purely low energy perspective in \cite{Guerrieri:2018uew, Guerrieri:2020bto}, both in the massive and massless cases,\footnote{See also \cite{Albert:2022oes,Fernandez:2022kzi,Albert:2023jtd,Ma:2023vgc} for applications of EFT bootstrap on pions at large $N_c$.} by including a minimal set of phenomenological parameters such as resonance masses and/or scattering lengths. The question still remained of how to incorporate high energy data into the bootstrap in order to identify the low energy theory corresponding to a given high energy one. 
  
 In a very important development in the S-matrix bootstrap program, Karateev, Kuhn and Penedones \cite{Karateev:2019ymz} proposed to use form factors and their related spectral densities to connect the low energy bootstrap with the UV theory. For example this allows to bound the central charge of the UV theory  as also recently done in \cite{Cordova:2023wjp} for $2d$ $O(N)$ models. More generally, this framework allows to incorporate UV information into the bootstrap as demonstrated in \cite{Chen:2021pgx,Correia:2022dyp} for certain 2d theories. 

 In this paper we have to deal with the case where there is spontaneous symmetry breaking which introduces a further complication, namely that the vacuum of the UV theory and the IR theory are not the same.  
  In the IR vacuum several operators other than the identity have expectation values and those expectation values are extra IR parameters. This important observation is one of the basis of the famous SVZ (or ITEP) sum rules which were introduced as a way to parameterize non-perturbative QCD physics in terms of a few parameters such as the quark and gluon condensates. See \cite{Novikov:1977dq,Reinders:1981bq,Reinders:1984sr} for reviews at that time and \cite{Gubler:2018ctz} for a recent survey, the original papers are \cite{SHIFMAN1979385,SHIFMAN1979448,Shifman:1978bw}. The SVZ sum rules have been extensively used in QCD phenomenology to do meson spectroscopy, specially for heavy quarks but also for glueballs \cite{Caron-Huot:2023tpw}. Light quarks can also be addressed, for example, already in the original paper \cite{SHIFMAN1979448} the mass of the $\rho$ meson was derived by assuming that it is a narrow resonance that saturates the spectral density. They can also be used to study spontaneous symmetry breaking in $\lambda\phi^4$ theory \cite{PhysRevD.25.838,PhysRevD.28.1364}. These and many other subsequent phenomenological results \cite{Gubler:2018ctz} suggest that this procedure may indeed lead to good results for the partial waves when incorporated into the S-matrix bootstrap framework in the case of spontaneous symmetry breaking and, in particular, for gauge theories.   
  \begin{figure}[t]
  	\centering
  	\includegraphics[width=0.75\textwidth]{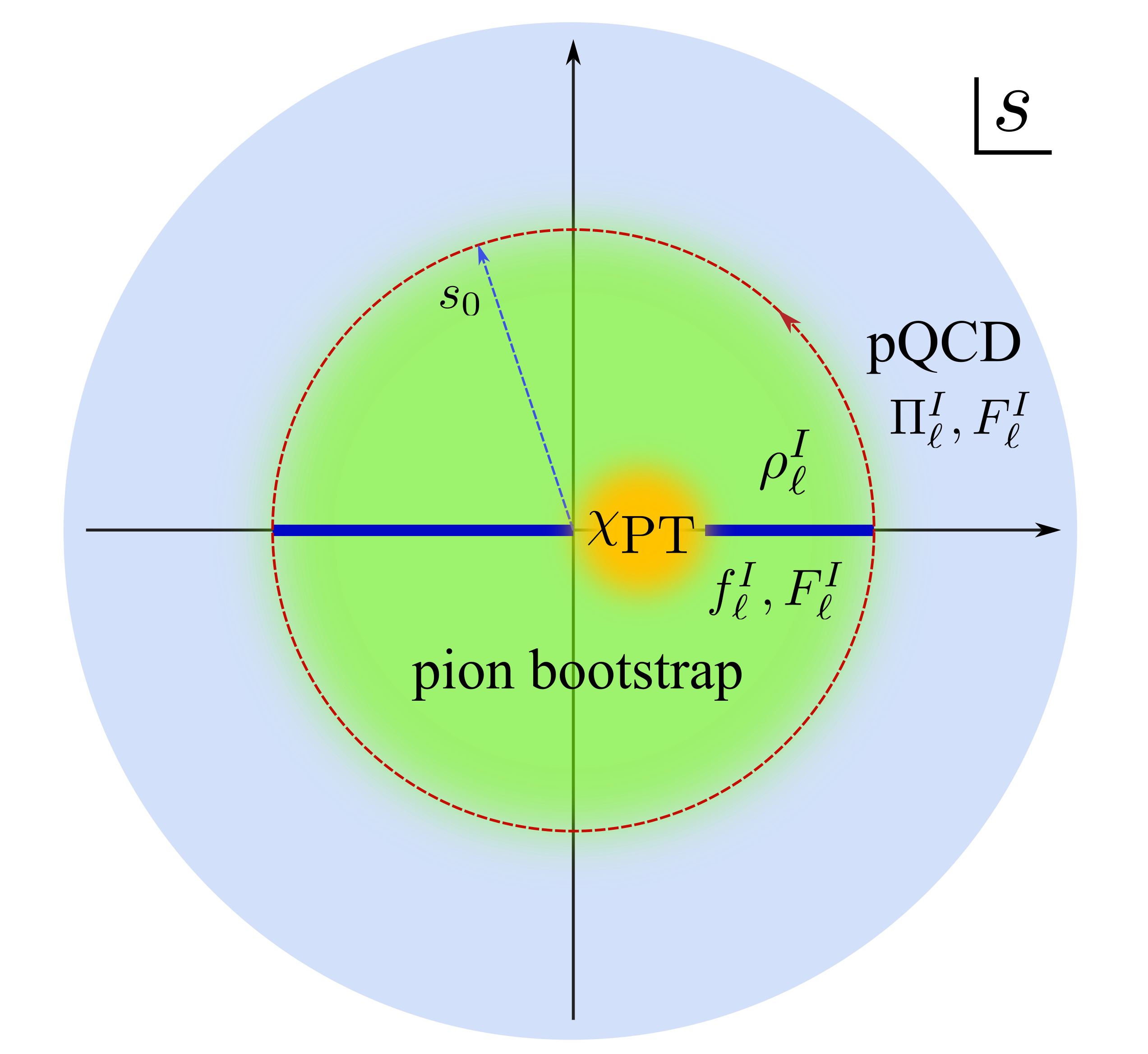}
  	\caption{In the $s$ complex plane we have analytic functions with possible cuts on the real axis: the partial waves $f^I_\ell(s)$, the form factors $F^I_\ell(s)$ and the two current correlators  $\Pi^I_\ell(s)$. The gauge theory bootstrap matches those functions across different scales described by chiral perturbation theory ($\chi_{\mbox{PT}}$) at low energy, the S-matrix/form factor bootstrap at intermediate energies where pion scattering is expected to saturate unitarity and perturbative QCD (pQCD) at large energies. We take $s_0$ such that $\alpha_s\simeq 0.4$.}
  	\label{method}
  \end{figure}
  Therefore, in this paper we combine the two approaches to write a bootstrap problem for the IR theory using UV properties as data together with IR observables such as the quark and gluon condensates that can be obtained from lattice computations\footnote{It might also be interesting to use the condensates as bootstrap parameters to find allowed values, for example.}.  
  \bigskip
  
  {\bf\underline{Main idea of the method:} } Our approach is pictorially summarized in fig.\ref{method} and can be described as follows:
  the high energy form of the current two point function $\Pi^{I}_\ell$ is obtained from perturbative QCD together with some IR information given by the vacuum condensates. Using finite energy sum rules one can relate that behavior to the spectral density $\rho^I_{\ell}$ in the intermediate energy region. In that region the spectral density of the two current correlator is saturated by two pion states and therefore given by the modulus squared of the current form factor $F^I_{\ell}$ (this is no longer true at high energies where the form factor goes to zero as predicted by pQCD). By analyticity, the modulus of the form factor is related to its phase. Since we also expect unitarity to be saturated, by Watson's theorem, the phase of the form-factor and partial waves $f^I_{\ell}$ are the same. In that way, the asymptotic form of the two current correlator given by the gauge theory trickles down to the partial waves.  
  
  On the other hand, in the unphysical region $0<s<4m_{\pi}^2$ the behavior of the partial waves is predicted by chiral  symmetry breaking which, together with the high energy information seems to fix their behavior.

\bigskip  
  This framework seems quite natural and it is plausible that it can lead to the full S-matrix in the infrared if enough UV operators are incorporated. 
  
  To test our approach, in this paper we apply it to the interesting case of $N_c=3$ and $N_f=2$ and obtain results that compare well with (QCD) pion physics as seen in figs.\ref{P1phaseshift} and \ref{S0S2phaseshift}. We should emphasize that, other than the QCD parameters $N_c$ (number of colors), $m_q$ (quark mass) and $\lqcd$ (QCD scale), the only numerical parameters we need to know are the pion mass $m_\pi$ and the pion decay constant $f_\pi$. The quark condensate can be determined from those by using the GMOR\cite{PhysRev.175.2195} relation and the gluon condensate is taken from the literature but they do not really play a significant role at the precision we work with. It seems clear that by considering a large number of operators, more precise results should be obtained. Although QCD is a primary motivation for this work, we emphasize that the main objective of this paper is to describe a framework to understand general gauge theories at low energy. Other theories can be of interest since lattice results can in principle compute certain phase shifts and then be used to compare with the bootstrap. 
    
\section{Low energy physics and the bootstrap}

 Consider an $SU(N_c)$ asymptotically free gauge theory with $N_f$ massive quarks with the same mass $m_q$ in the fundamental representation of $SU(N_c)$. Assuming chiral symmetry breaking 
 \begin{equation}\label{h1}
SU(N_f)_L\times SU(N_f)_R\to SU(N_f)_V
 \end{equation} 
 and confinement with $m_q\ll \lqcd$, the low energy physics is dominated by the pseudo-Goldstone bosons whose low energy Lagrangian is given by the chiral Lagrangian. One of the basic objective of the S-matrix bootstrap program (see \eg\ \cite{Kruczenski:2022lot}) is to compute the S-matrix of the pseudo-Goldstone bosons. In the chiral limit of massless quarks, the pion would be massless but here there is a small amount of explicit chiral symmetry breaking due to the small quark mass and therefore the pions are massive, $m_\pi\neq 0$. In this paper we measure everything using the pion mass as unit so that
 \beq\label{h2}
  m_\pi=1
 \eeq    
Other dimensionful quantities are reduced to this units using a value of $m_\pi=140\,\MeV$. We restore units to compare with experiments, for example when plotting phase-shifts. 

 Finally, besides the pion mass, the input to the calculation should be the high energy theory parameterized by $N_c$, $N_f$, $m_q$ and $\lqcd$ together with (as few as possible) low energy quantities that can be computed for example using lattice gauge theory. In this paper we use $f_\pi=92\,\MeV$, the pion decay constant, the quark condensate and the gluon condensate. 
 
 \medskip
 
 With all this in mind, the steps of the gauge theory bootstrap are:
 \begin{itemize}
 	\item{\bf Pure S-matrix bootstrap:} In the first step the bootstrap method should map out the space of low-energy S-matrices allowed by unitarity, crossing, analyticity and the global $SU(N_f)_V$ flavor symmetry.
 	\item{\bf Chiral symmetry breaking:} In the second step we add the constraints of chiral perturbation theory that determines the low energy S-matrix in the unphysical region (inside the Madelstam triangle) in terms of a few, in principle unknown, parameters. In this paper we  use a linear approximation (Weinberg's model) and the values of $m_\pi$, $f_\pi$.   
 	\item{\bf pQCD, form factors and SVZ:}  Finally, we pick a scale $s_0$ where the strong coupling constant has what we consider an adequate value to match high energy and low energy (here we pick $\alpha_s\simeq 0.4$ corresponding to $s_0\simeq 1.2$GeV).
 	   	 The information on the high energy theory is introduced for each partial wave (here only $S0$ and $P1$)  via form factors and the SVZ sum rules. In terms of the UV fixed point (in this case a free theory), we are using the spectrum of CFT operators and the OPE expansion of two operators in terms of operators whose expectation in the QCD vacuum is not zero. See eq. \eqref{svzexp} below.
 \end{itemize}
 To be concrete, and also due to available QCD experimental data, in this paper, we consider only the case of $N_f=2$ and $N_c=3$. Notice that different values of $N_f$ correspond to different low energy global symmetries and therefore to different S-matrix bootstrap setups, but $N_c$ is just a parameter that appears in the SVZ sum rules. Although we compare some results to experiment (figs.\ref{P1phaseshift} and \ref{S0S2phaseshift}), the objective of this paper is not to do phenomenology but to {\it provide a method that can be used to study the low energy physics of generic gauge theories}.
 
\subsection{Pion bootstrap with analyticity, crossing, unitarity and $SU(2)_V$ symmetry}         
  
  In the case of $N_f=2$, the low energy physics\footnote{See \eg\ \cite{donoghue_golowich_holstein_2014}} is described in terms of pions parameterizing an $S^3$ sphere. The scattering amplitude for the $2\rightarrow 2$ scattering
\beq
\pi_a(p_1) + \pi_b(p_2) \rightarrow \pi_c(p_3) +\pi_d(p_4), 
\label{b2}
\eeq
is given by an analytic function $A(s,t,u)$ of the Mandelstam variables $s,t,u$ as
\beq
T_{ab,cd} = A(s,t,u) \delta_{ab} \delta_{cd} + A(t,s,u) \delta_{ac} \delta_{bd} + A(u,t,s) \delta_{ad}\delta_{bc} 
\label{a8}
\eeq
with $A(s,t,u)=A(s,u,t)$ and $a,b,c,d=1,2,3$. The crossing and isospin symmetries are manifest. To impose the unitarity constraint we consider amplitudes of well-defined isospin ($I=0,1,2$) in the $s$-channel:
\beq
T_{ab,cd} = \frac{1}{3} T^{I=0} \delta_{ab}\delta_{cd} + \frac{1}{2} T^{I=1} (\delta_{ac}\delta_{bd}-\delta_{ad}\delta_{bc}) + \half T^{I=2} (\delta_{ac}\delta_{bd}+\delta_{ad}\delta_{bc}-\frac{2}{3}\delta_{ab}\delta_{cd})  
\label{a9}
\eeq 
with 
\begin{subequations}\label{a10}
	\beqa	
	T^{I=0}(s,t,u) &=& 3 A(s,t,u)+A(t,s,u)+A(u,t,s) \\ 
	T^{I=1}(s,t,u) &=& A(t,s,u)-A(u,t,s) \\ 
	T^{I=2}(s,t,u) &=& A(t,s,u)+A(u,t,s) 
	\eeqa
\end{subequations}
We also use the Mandelstam representation for the scattering amplitude, given by\footnote{It is customary to denote the double spectral density in the Mandelstam representation by $\rho_\alpha(x,y)$. It should not be confused with the spectral density of the current correlators denoted $\rho^I_\ell(x)$ later in the paper.}
\beqa\label{Adef}
A(s,t,u) &=& T_0 + \frac{1}{\pi} \intx  \frac{\sigma_1(x)}{x-s}+\frac{1}{\pi} \intx \sigma_2(x) \left[\frac{1}{x-t}+\frac{1}{x-u}\right]\nonumber \\ 
         && +  \frac{1}{\pi^2}  \intxy  \frac{\rho_1(x,y)}{x-s}\left[\frac{1}{y-t}+\frac{1}{y-u}\right] \nonumber\\
         && +  \frac{1}{\pi^2}  \intxy   \frac{\rho_2(x,y)}{(x-t)(y-u)}
\label{a11}
\eeqa
with $\rho_2(x,y)=\rho_2(y,x)$. As mentioned before we are setting $m_\pi=1$. The scattering amplitude is then parametrized by the variables
\begin{equation}\label{Apara}
	T_0,\;\; \sigma_{\a=1,2}(x),\;\; \rho_{\a=1,2}(x,y)
\end{equation}

To impose the unitarity constraint, we consider the  partial waves:
\beq\label{fdef}
f^I_\ell(s) = \frac{1}{4} \intP T^I(s,t)
\eeq
with
\beq
	t = -\frac{(s-4)(1-\mu)}{2} \label{a4}	
\eeq
and $u=4-s-t$. The even $\ell$'s are non-vanishing for $I=0,2$ and the odd ones are non-vanishing for $I=1$. For later use, notice that \eqref{fdef}, \eqref{a4} also allow to define the partial waves for complex values of $s$ if we have the analytic functions $T^I(s,t)$. The partial scattering amplitude for a given $I,\ell$ is
\begin{equation}\label{h4}
	S_\ell^I(s^+)=1+i\pi\sqrt{\frac{s-4}{s}}f^I_\ell(s)=\eta^I_\ell(s)\, e^{2i\delta^I_{\ell}(s)},\;\; s\in\mathbb{R}_{\ge 4}
\end{equation}
where $\delta^{I}_\ell$ are the phase shifts and $\eta^I_\ell=|S^I_\ell(s)|$ satisfies the unitarity constraint:
\beq\label{uni}
 \eta^I_\ell(s) = |S^I_\ell(s)| \le 1,\;\;\; s\in\mathbb{R}_{\ge 4},\; \; \forall \ell, I, 
\eeq 
In bootstrap calculations one typically obtains that unitarity is saturated $\eta^I_\ell(s)=1$. This is usually a problem since interacting theories do not saturate unitarity at all energies \cite{Aks:1965qga}.\footnote{See {\it e.g.} \cite{Antunes:2023irg,Tourkine:2023xtu} for recent work exploring inelasticity in the S-matrix bootstrap.} However here we are using the bootstrap in the region $s<s_0=1.2\,\GeV$ where unitarity is approximately saturated in QCD.\footnote{As seen, for example, from the experimental data in \cite{Protopopescu:1973sh,LOSTY1974185,HYAMS1973134}.}
This is the basic bootstrap setup that allows numerical methods to be used to map out the space of 2-to-2 S-matrices that satisfy all constraints. This is step one. Step two is to impose the constraints from chiral symmetry breaking that we do in the next section. 

\subsection{Chiral symmetry breaking}\label{chibreak}

Chiral symmetry breaking not only requires the existence of pseudo-Goldstone boson, i.e., the pions, but also predicts their low energy properties. Parameterizing the pion fields with
\begin{equation}
U=e^{i\frac{\pi_a\tau_a}{f_\pi}}
\end{equation}
where $\tau_a, a=1,2,3$ are the Pauli matrices, the effective lagrangian, at lowest order in the derivative expansion can be written as \cite{GASSER198765,GASSER1984142,donoghue_golowich_holstein_2014}
\beq
 \cL = \frac{f_\pi^2}{4}\left\{\tr\left(\partial_\mu U \partial^\mu U^\dagger\right)+ m_\pi^2\tr\left(U+U^\dagger\right)\right\} 
\eeq
or, expanding in powers of the pion field
\beqa 
 \cL_2^{2\pi}   &=& \half \partial_\mu \pi_a \partial^\mu \pi_a - \half m_\pi^2 \pi_a \pi_a  \\ 
 \cL_2^{4\pi}   &=& \frac{1}{6f_\pi^2}\left(\pi_a\partial_\mu\pi_a\pi_b\partial^\mu\pi_b-\pi_a\pi_a\partial_\mu\pi_b\partial^\mu\pi_b\right)+\frac{m_\pi^2}{24f_\pi^2}\pi_a\pi_a\pi_b\pi_b
\eeqa
At the lowest order, one has Weinberg's model that can be obtained at tree level from the interactions in $\cL_2^{4\pi}$. The amplitude is given by \cite{PhysRevLett.17.616,GASSER198765,GASSER1984142,donoghue_golowich_holstein_2014}:\footnote{Our definition of the amplitudes \eqref{a8} differ by a factor of $8\pi^2$ from the definition of \eg\ \cite{Peskin:1995ev}: $T_{ab,cd}=\frac{1}{8\pi^2}\cM(\pi^a\pi^b\to\pi^c\pi^d)$.}
\beq\label{h5}
 A(s,t,u) =  \frac{s-m_{\pi}^2}{8\pi^2\,f_\pi^2}
\eeq
from where the leading behavior of the pion partial waves follows (see \eg\ \cite{donoghue_golowich_holstein_2014}):
\begin{equation}\label{chiralratio}
f^0_0(s)=\frac{2}{\pi}\frac{2s-m_{\pi}^2}{32\pi f_{\pi}^2},\;\; f^1_1(s)=\frac{2}{\pi}\frac{s-4m_{\pi}^2}{96\pi f_{\pi}^2},\;\; f^2_0(s)=\frac{2}{\pi}\frac{2m_{\pi}^2-s}{32\pi f_{\pi}^2}
\end{equation}
where we have temporarily restored the pion mass $m_{\pi}$ in the expressions.
Matching with the phase shifts at threshold this implies, for increasing $s\ge 4$, a rapidly growing $\delta_{\ell=0}^{I=0}$ phase shift and a slowly falling $\delta_{\ell=0}^{I=2}$, the signature of the Weinberg model. We expect these low energy properties to remain true as long as we are in the neighborhood of such model even if the partial waves are not exactly linear in the unphysical region.      
The pion decay constant $f_\pi$ determines the effective coupling of pions as $\lambda_{\text{eff}} \sim \frac{s}{f_\pi^2}$. For the linearized approximation to be reasonably valid within the unphysical region we need that the corrections controlled by $\lambda_{\text{eff}}$ are small up to $s\sim 4 m_\pi^2$ which means that $f_\pi/m_\pi$ is bounded from below in this setup. Notice that, in the linearized approximation, the amplitude \eqref{a8} for $\pi^0+\pi^0\rightarrow \pi^0+\pi^0$ at the symmetric point is given by the value:
\beq
\begin{aligned}
	\lambda\equiv&\frac{1}{32\pi}\cM(\pi^0\pi^0\to\pi^0\pi^0)|_{s=t=u=4/3}\\
	 =& \frac{\pi}{4} T_{33,33}\Big(\frac{4}{3},\frac{4}{3},\frac{4}{3}\Big)=\frac{m_\pi^2}{32\pi{f_\pi}^2}\simeq 0.023\ll \lambda_{\text{max}}\simeq 2.661
\end{aligned}
\eeq
 where $\lambda_{\text{max}}$ is the largest value allowed by analyticity, crossing and unitarity \cite{Paulos:2017fhb} in the normalization conventional when studying amplitude bounds. \footnote{The same value was obtained using the dual approach in \cite{He:2021eqn}.}

\smallskip

The next step is to use form factors and the SVZ sum rules to incorporate information about the UV theory. 

\subsection{Form factor bootstrap and SVZ sum rules}

   The form factor bootstrap was introduced in \cite{Karateev:2019ymz}, both, as a way to compute form factors and to introduce information on the UV fixed point into the low energy bootstrap. In this section we review those results and extend them to include the SVZ sum rules. As mentioned, we expect these sum rules to provide important information since one of the original purposes of the SVZ program was to do hadron spectroscopy and as such they have extensively been used in QCD phenomenology (see \eg\ \cite{donoghue_golowich_holstein_2014, Peskin:1995ev,  narison_2004} and references therein). A recent example is \cite{cherry2001qcd} where the scalar sum rules are discussed in detail.

   \subsubsection{Form factors}
   
   Since all the main derivations are known,\footnote{See \cite{Karateev:2020axc} and references therein for detailed analysis on the two-point functions and form factors.} we try to be brief, emphasize the normalization, and mainly present the results that are used in the numerical section.  We start with single particle states normalized as\footnote{We mostly follow \cite{Peskin:1995ev} for normalizations and basic computations.}
   \beq\label{h6}
   \bra{p}p'\rangle = 2p_0 (2\pi)^3 \delta^{(3)}(\vec{p}-\vec{p}')
   \eeq
   \beq\label{h7}
   \mathbb{1} = \int\frac{d^3\vec{p}}{2p_0 (2\pi)^3}\ \ket{p}\bra{p} 
   \eeq
   For two (non-identical) particle states we also consider the states $\ket{P,\ell\sigma}$ with center of mass momentum $P$ and angular momentum $\ell$ and polarization $\sigma$ defined as
   \beq\label{h8}
   \bra{p_1, p_2} P,\ell,\sigma\rangle = 16\pi^3 \sqrt{\frac{P_0}{|\vec{p}_1|}}\ \delta^{(4)}(P-p_1-p_2)\, Y_{\ell\sigma}(\hat{p}_1)
   \eeq
   where $Y_{\ell\sigma}$ are the standard spherical harmonics. These states are normalized to
\beq\label{h9}
 \bra{P',\ell'\sigma'}P,\ell\sigma\rangle = \delta^{(4)}(P'-P) \delta_{\ell\ell'}\delta_{\sigma\sigma'} 
\eeq
   In most cases we are going to work in the center of mass frame where $P=(P_0,0,0,0)$.   
   
   For each partial wave with isospin $I$ and angular momentum $\ell$, we consider the operator $\cO_\ell^I$ of the higher energy theory that has the same quantum numbers and the lowest conformal dimension. For example, using the standard $u,d$ quark notation, we have
   \begin{subequations}
   	   \beqa
   	S0&:&\ j_S(x)= m_q (\bar{u}u+\bar{d}d)\label{js} \\
   	P1&:&\ j^\mu_V(x) = \half (\bar{u}\gamma^\mu u-\bar{d}\gamma^\mu d)\label{jv}
   	\eeqa 
   \end{subequations}
   where $S0$ and $P1$ stand for $I=0,\ell=0$ and $I=1,\ell=1$ respectively.
  For higher partial waves with $I=0,1$, we can take 
\beq\label{h10}
  j^{I}_{\ell,\Delta} = \Delta_{\mu_1}\ldots\Delta_{\mu_\ell} t^I_{ab}\ (\bar{q}_a  \gamma^{\mu_1} D^{\mu_2}\ldots D^{\mu_\ell} q_b) 
\eeq
where $\Delta_\mu$ is a constant vector such that $\Delta^2=0$ so that only the traceless symmetric part contributes.  The matrix $t^I_{ab}$ composes the isospin to $I=0,1$ and the covariant derivatives $D^\mu$ are required to make the operator gauge invariant. For $I=2$ we need a four quark operator such as 
\beq\label{h11}
j^{I=2}_{\ell=0}=(\bar{d}u)(\bar{d}u).
\eeq 
In any case we leave those higher partial waves for future work. 

\smallskip

Going back to the $S0$, $P1$ waves, we notice that the vector current \eqref{jv} is conserved $\partial_\mu j^\mu_V=0$.  Taking into account this conservation law and Lorentz symmetry we can write the pion form factors in standard form
  \begin{subequations}
  \beqa
     {}_{\text{out}}\bra{\pi^+(p_2)} j_S(x) \ket{\pi^+(p_1)}_{\text{in}} &=& e^{i(p_1-p_2)x}\  F_0(t) \label{h13}\\
     {}_{\text{out}}\bra{\pi^+(p_2)} j^\mu_V(x) \ket{\pi^+(p_1)}_{\text{in}} &=& e^{i(p_1-p_2)x}\  (p_1^\mu+p_2^\mu) \, F_1(t) \label{h14}
  \eeqa  
  \end{subequations}
   where $t=(p_1-p_2)^2$. The labels in and out are useful but redundant for one particle states. Using crossing symmetry (and replacing $p_1\rightarrow -p_1$) we obtain
   \beq\label{F1def}
   {}_{\text{out}}\bra{\pi^-(p_1)\ \pi^+(p_2)} j^\mu_V(0) \ket{0} =  (p_2^\mu-p_1^\mu) \, F_1(s=(p_1+p_2)^2) 
   \eeq
   and also 
   \beq\label{h15}
   \bra{0} j^\mu_V(0) \ket{\pi^+(p_1)\ \pi^-(p_2)}_{\text{in}} = (p_1^\mu-p_2^\mu) \, F_1(s) 
   \eeq
   or, equivalently
   \beq\label{h16}
   {}_{\text{in}}\bra{\pi^-(p_1)\ \pi^+(p_2)} j^\mu_V(0) \ket{0} =  (p_2^\mu-p_1^\mu) \, F^*_1(s) 
   \eeq
   The same can be done with the scalar form factor, \ie\ for the operator \eqref{js}:
   \begin{subequations}
   \beqa
   {}_{\text{out}}\bra{\pi^-(p_1)\ \pi^+(p_2)} j_S(0) \ket{0}  &=&   \, F_0(s)   \label{F0def}\\
   {}_{\text{in}}\bra{\pi^-(p_1)\  \pi^+(p_2)} j_S(0) \ket{0}  &=&   \, F^*_0(s)  \label{h17}
   \eeqa
\end{subequations}
  For states of fixed angular momentum and isospin we find
  \begin{subequations}
     \beqa
  {}_{\text{out}}\bra{I=0, I_3=0; P\ell\sigma } j_S(0) \ket{0} &=&   \cF^0_0(s)  \delta_{\ell0}\delta_{\sigma 0} \label{h18}\\
  {}_{\text{out}}\bra{I=1, I_3=0; P\ell\sigma } j_{V,\sigma'}(0) \ket{0} &=&    \cF^1_1(s)  \delta_{\ell 1} \delta_{\sigma\sigma'} \label{F1cur}
  \eeqa
  \end{subequations}
   with
   \begin{subequations}\label{curF}
      \beqa
   \cF^0_0(s)  &=&  \frac{\sqrt{6\pi}}{16\pi^3}\frac{1}{s^{\frac{1}{4}}}\left(\frac{s-4}{4}\right)^{\frac{1}{4}}\, F_0(s)   \label{h19} \\
   \cF^1_1(s)  &=&  \sqrt{\frac{4\pi}{3}}\frac{1}{8\pi^3}\frac{1}{s^{\frac{1}{4}}}\left(\frac{s-4}{4}\right)^{\frac{3}{4}}\, F_1(s) \label{h20}
   \eeqa
   \end{subequations}
  and in \eqref{F1cur} we defined
  \beq
   j_{V,0}=j_{V,z}, \ j_{V,\pm}=\mp\frac{1}{\sqrt{2}} (j_{V,x}\pm i j_{V,y})\ . \label{h21}
  \eeq
   Here $j_{V,0}=j_{V,z}$ is not to be confused with the time component $j^{\mu=0}_V$ whose form factor vanishes.  Now, using that $Q_V=\int d^3 x\, j^0_V(x)=\half(n_u-n_d)$ we obtain
   \beq\label{F1norm}
   F_1(0)=1\ ,
   \eeq
   where we used that the $\pi^+$ quark content is $u\bar{d}$. For the scalar current we know \cite{DONOGHUE1990341,Ananthanarayan_2004}:
   \beq\label{F0norm}
    F_0(0)=m_q\frac{\partial m_\pi^2}{\partial m_q} \simeq m_\pi^2 = 1\ ,
   \eeq
   where the last equality is just that we use $m_\pi$ as the unit of mass.  The way to derive this is to realize that $j_S = -m_q\frac{\delta \cL_{\mbox{\scriptsize QCD}}}{\delta m_q}$ where $\cL_{\mbox{\scriptsize QCD}}$ is the QCD Lagrangian. 
   	In the low energy effective theory $-m_q\frac{\delta \cL_{\mbox{\scriptsize eff}}}{\delta m_q}=\half m_q \frac{\partial m_{\pi}^2}{\partial m_q} \pi_a^2=j_S$. Computing the form factor using the low energy form of the current gives \eqref{F0norm}. 
   	
   	Given that $F_{0,1}(s=0)=1$ and that the form factors are analytic functions of $s$ with a cut on the real axis for $s>4$ we can write a subtracted dispersion relation 
   	\begin{equation}\label{Fdisp}
   	F_{\ell}(s)=1+\frac{1}{\pi}\int_4^{\infty}\!\!\! dx\bigg(\frac{1}{x-s}-\frac{1}{x}\bigg)\text{Im}F_{\ell}(x)
   	\end{equation}
   that we use in the numerical section to parameterize the form factor in terms of its discontinuity $\text{Im}F_{\ell}(x)$ across the cut. 
   	
   \subsubsection{Current correlators and bootstrap}
   Finally we introduce the vacuum polarizations (a.k.a. current correlators)
   \begin{subequations}
   	   \beqa
   	\Pi^0_0(s) &=& i\int \frac{d^4x}{(2\pi)^4} e^{iPx} \bra{0} \hat{T}\left\{j_S(x)j_S(0)\right\}\ket{0} \label{cc1}\\ 
   	\Pi^1_1(s) \delta_{\sigma'\sigma} &=& i\int \frac{d^4x}{(2\pi)^4} e^{iPx} \bra{0} \hat{T}\left\{j_{V,\sigma'}^\dagger(x)j_{V,\sigma}(0)\right\}\ket{0} \label{cc2}
   	\eeqa
   \end{subequations}
   These are analytic functions of $s$ with a cut on the real axis for $s>4$. The spectral density, {\it i.e.,} discontinuity along the cut is given by
   \beq\label{h22}
     \rho^I_\ell(s) = 2\, \Im \Pi^I_\ell(x+i\epsilon) = \int\frac{d^4 x}{(2\pi)^4} e^{iPx} \bra{0} j_\ell^{I \dagger}(x) j^I_\ell(0)\ket{0}
   \eeq
   
   \smallskip
   
 After all these preliminary definitions we are ready to introduce the KKP form factor bootstrap \cite{Karateev:2019ymz}. We can do $I,\ell=0,1$ simultaneously by considering $j^0_0=j_S$ and $j^1_1=j_{V,+}$.
 We define an operator 
 \beq\label{h23}
  \cO_{P,I,\ell} = \int \frac{d^4x}{(2\pi)^4} e^{-iPx} j^I_\ell(x)
 \eeq
 and consider the positive semi-definite matrix of overlaps
   \beq\label{Bmt}
   \begin{array}{c c} &
   	\begin{array}{c c c} \ket{\mbox{out}}_{P,I,\ell} & \ket{\mbox{in}}_{P,I,\ell} & \cO_{P,I,\ell}\ket{0} \\
   	\end{array}
   	\\
   	\begin{array}{c c c}
   		\bra{\mbox{out}}_{P',I,\ell}  \\
   		\bra{\mbox{in}}_{P',I,\ell}\\
   		\bra{0} \cO^\dagger_{P',I,\ell}
   	\end{array}
   	&
   	\left(
   	\begin{array}{c c c}
   		1 &\ S^I_\ell(s)\ &\ \ \cF^I_\ell \\
   		S_\ell^{I*}(s) &\ 1\ & \ \ \cF^{I*}_\ell \\
   		\cF^{I*}_\ell &\ \cF^I_\ell\ &\ \ \rho^I_\ell(s)
   	\end{array}
   	\right) \succeq 0
   \end{array} 
   \eeq
   where we removed an overall factor $\delta^4(P-P')$, namely the identity in the center of mass variable.  
   In particular, this implies the constraints
   \begin{subequations}\label{Fcurbound}
     \beqa
  |\cF^0_0(s)|^2 &=& \frac{1}{(2\pi)^4} \frac{3}{16\pi}\sqrt{\frac{s-4}{s}}\, |F_0(s)|^2 \le \rho^0_0(s) \label{h24}\\
  |\cF^1_1(s)|^2 &=& \frac{1}{(2\pi)^4} \frac{1}{24\pi} \frac{(s-4)^{\frac{3}{2}}}{\sqrt{s}}\, |F_1(s)|^2 \le \rho^1_1(s) \label{h25}
  \eeqa
   \end{subequations}
If this conditions are saturated it means that the spectral density is saturated by two pion states. At high energy, instead, it is saturated by multiparticle states and the form factors are much lower than this bound.  
To summarize, the form-factor bootstrap is
parametrized by the variables
\begin{equation}\label{ffApara}
T_0,\;\; \sigma_{\a=1,2}(x),\;\; \rho_{\a=1,2}(x,y),\;\; \text{Im}F_{\ell}(x), \;\; \rho^I_{\ell}(x).
\end{equation} 
subject to the constraints \eqref{Bmt}.

\subsubsection{SVZ expansion and FESR}   
   
In the limit of large energy we have that, at leading order in $s\rightarrow \infty$ and leading order in perturbation theory, the current correlators \eqref{cc1}, \eqref{cc2} behave as:
   \begin{subequations}
   	\beqa
   	\Pi^0_0(s) &\simeq& \frac{N_c N_f m_q^2}{(2\pi)^4}\  \frac{(-s)}{8\pi^2}\ln(-\frac{s}{\mu^2})  \label{h26} \\
   	\Pi^1_1(s) &\simeq& \frac{N_c}{(2\pi)^4}\frac{(-s)}{24\pi^2}\ln(-\frac{s}{\mu^2})               \label{h27}
   	\eeqa
   \end{subequations}
which allows to introduce $N_c$ in the low energy theory and should give a reasonable initial approximation for the bootstrap. Here, $\mu$ is the renormalization scale that we can take as $\mu^2=s_0$, the scale at which we later match with the bootstrap.  
   The previous result just computes the current correlator in the high energy vacuum using a quark loop and can be interpreted as keeping the component proportional to the identity in the current OPE:
\beq
  T\{j(x) j(0)\} = C_{\mathbb{1}}(x)\ \mathbb{1}  + \sum_\cO C_{\cO}(x)\ \cO(0)   \label{h28}
\eeq   
However, in the symmetry broken vacuum, other operators contribute and we get the SVZ expansion
 \beq\label{svzexp}
\bra{0} T\{j(x) j(0)\} \ket{0} = C_{\mathbb{1}}(x)  +  C_{\bar{q} q}(x)\ \bra{0}j_S(0)\ket{0} + C_{G^2}(x)\ \bra{0} \frac{\alpha_s}{\pi} G_{\mu\nu}^a G^{a\,\mu\nu} \ket{0} + \ldots
 \eeq   
where $j_S=m_q(\bar{u}u+\bar{d}d)$, as in \eqref{js}.
In addition one can also compute the subleading order in perturbation theory. Altogether, for the case of $N_c=3$ it reads \cite{SHIFMAN1979385,Reinders:1981bq} 
   \beqa
    \Pi^0_0(s) &\simeq& \frac{N_f m_q^2}{(2\pi)^4}  \left\{-\frac{3}{8\pi^2}\left(1+\frac{13}{3}\frac{\alpha_s}{\pi}\right)s\ln(-\frac{s}{\mu^2}) - \frac{1}{8s} \langle\frac{\alpha_s}{\pi}G^2\rangle - \frac{3}{2s}\langle j_S\rangle+\ldots \right\} \non\\
   \Pi^1_1(s) &\simeq& \half\frac{1}{(2\pi)^4}  \left\{-\frac{1}{4\pi^2}\left(1+\frac{\alpha_s}{\pi}\right)s\ln(-\frac{s}{\mu^2}) + \frac{1}{12s} \langle\frac{\alpha_s}{\pi}G^2\rangle + \frac{1}{s}\langle j_S\rangle +\ldots \right\} \non \\  \label{h29}
   \eeqa
The condensates $\langle\frac{\alpha_s}{\pi}G^2\rangle$, $\langle j_S\rangle $ are extra parameters of the bootstrap can be obtained from lattice calculations but are only required for high precision calculations or in the case of heavy quarks.

  \begin{figure}[t]
	\centering
	\includegraphics[width=0.7\textwidth]{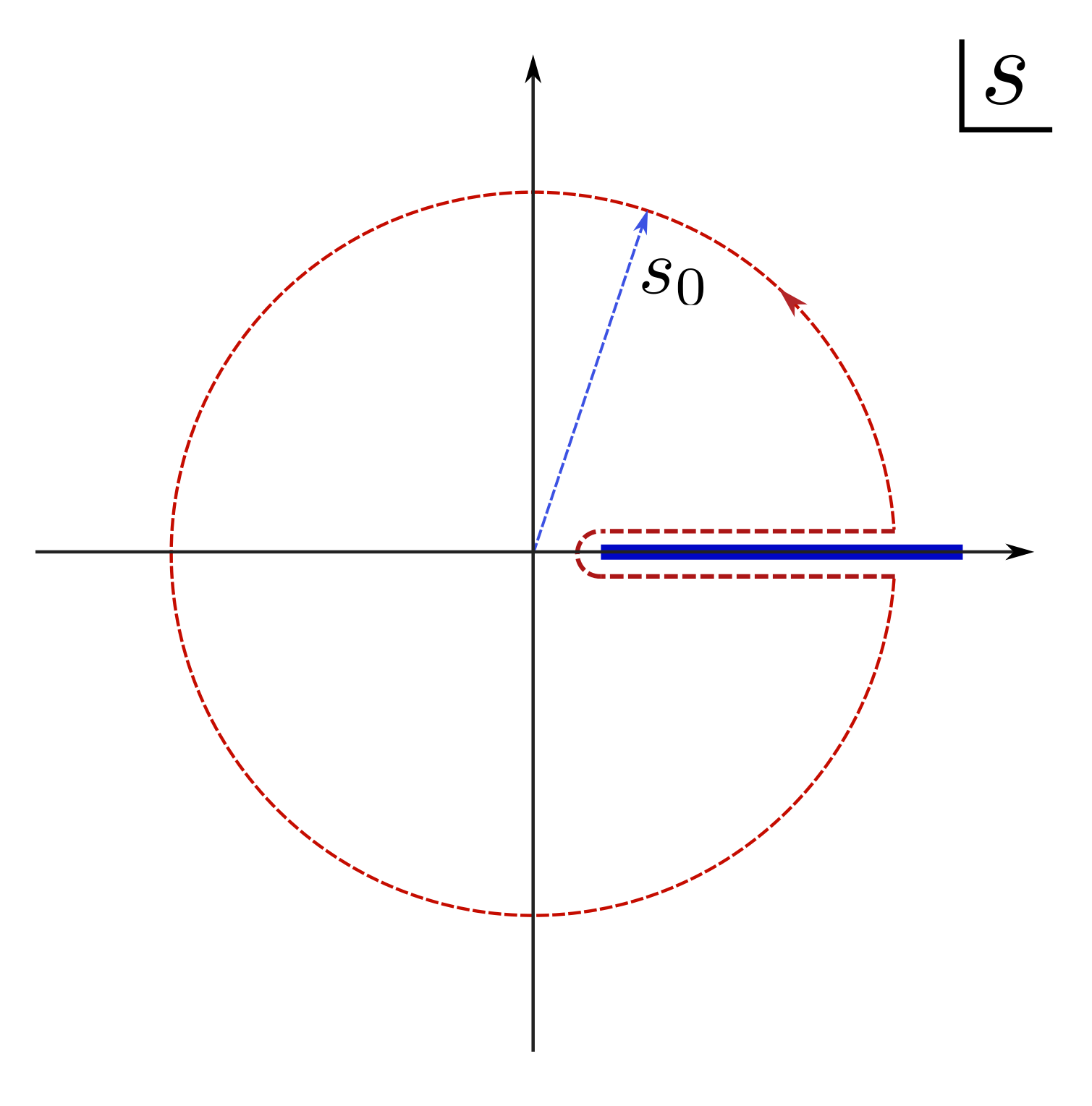}
	\caption{Contour of integration for the Finite Energy Sum Rule (FESR)  of an analytic function with a cut along the blue line. The integral around the red contour is zero. This relates the integral of the jump across the cut with the asymptotic behavior on the circle for large $s_0$.}
	\label{fesrfig}
\end{figure}
  
  Now we introduce the finite energy sum rules (FESR) in the standard manner (see \eg\ \cite{narison_2004}). The integral of an analytic function  around the contour illustrated in fig. \ref{fesrfig} vanishes. Here we consider $s^n \Pi^0_0(s)$, $n\in\mathbb{Z}_{\ge 0}$ and $s^n \Pi^1_1(s)$, $n\in\mathbb{Z}_{\ge -1}$ where the value $n=-1$ is allowed since $\Pi^1_1(s)$ has a zero at $s=0$ in the way we defined it. The contour integral has a contribution from the (large) circle at $|s|=s_0$ and one from the jump across the cut. All in all we obtain
  \beq\label{h30}
   \int_4^{s_0} \rho(x) x^n dx = -s_0^{n+1} \int_0^{2\pi} e^{i(n+1)\varphi} \Pi(s_0e^{i\varphi}) d\varphi
  \eeq 
  Using the expansion \eqref{h29} in the RHS of \eqref{h30}\footnote{To use the expansion \eqref{h29} for the large circle at $s_0$, we have employed the Sugawara-Kanazawa theorem \cite{Sugawara:1961zz}: if an analytic function has singularities only on the real axis and is polynomially bounded at infinity, then it has the same limit in every direction at infinity.} and $(n\in \mathbb{Z})$
\beq
   \int_0^{2\pi} e^{in\varphi} \ln(e^{i\varphi}) = \frac{2\pi}{n},\ n\neq 0,\;\; \int_0^{2\pi} e^{in\varphi} d\varphi = 2\pi \delta_n ,
\eeq
we find
\begin{equation}\label{srexpression}
	\begin{aligned}
	\int_4^{s_0} \rho^0_0(x) x^n dx =&  \frac{s_0^{n+1} N_f m_q^2}{(2\pi)^4} \left\{\frac{3s_0}{4\pi(n+2)}\left(1+\frac{13}{3}\frac{\alpha_s}{\pi}\right)\right.\\
	& \left. +\delta_n \frac{\pi}{4s_0} \langle\frac{\alpha_s}{\pi}G^2\rangle +\delta_n \frac{3\pi}{s_0} \langle j_S\rangle+\ldots\right\},\ \ n\ge 0 \\
	\int_4^{s_0} \rho^1_1(x) x^n dx =& - \frac{s_0^{n+1}}{(2\pi)^4} \half \left\{-\frac{s_0}{2\pi(n+2)}\left(1+\frac{\alpha_s}{\pi}\right)\right.\\
	& \left. +\delta_n \frac{\pi}{6s_0} \langle\frac{\alpha_s}{\pi}G^2\rangle +\delta_n\frac{2\pi}{s_0} \langle j_S\rangle+\ldots\right\} ,\ \ n\ge -1
	\end{aligned}
\end{equation} 
 Eqs. \eqref{srexpression} are linear constraints on the spectral density $\rho^I_\ell(x)$ that incorporate information on the UV theory which we impose on the bootstrap. In this paper we impose two moments for each partial wave, $n=0,1$ for the $S0$ wave and $n=-1,0$ for the $P1$ wave.
 Notice that the renormalization scale $\mu$ in \eqref{h29} drops from these results. For the future one should do the same for more partial waves and consider more condensates. 

\subsubsection{Asymptotic behavior of the form factors}\label{FFasymsection}

 Finally we need the high energy behavior of the form factor from QCD. Using a QCD/parton model one can get that the pion form factors scale as \cite{Pire:1996bc}:
 \beq
  |F_\pi(s)| \sim \frac{|q|}{|s| R_\pi^2} \label{Fscaling}
 \eeq   
where $R_\pi$ is the radius of the pion $\sim\frac{1}{f_\pi}$ and $q$ is the charge associated with the form factor. The physical intuition of why the form factor is small for large negative $s$ is as follows: a pion is moving very fast to the right and is struck by an off-shell photon and we want the amplitude for the pion to end up moving backwards with opposite momentum. The amplitude for this process should be quite small since the pion will most likely be destroyed instead. In \cite{osti_1447331} the more precise behavior
\beq
 F_\pi(s) \simeq - \frac{16\pi \alpha_s(s) f_\pi^2}{s}  \label{ffQCD}
\eeq    
for $s\rightarrow-\infty$ was found for the electromagnetic form factor.  We are not aware of a similar calculation for the scalar form factor but we expect the general behavior \eqref{Fscaling} with the ratio between form factors given by the respective charges 
\beq\label{ffratioasymp}
 \frac{|F_0|}{|F_1|} \sim 2 m_q
\eeq
%

\bigskip

Let us summarize again how the method manages to incorporate gauge theory information into the low energy bootstrap. The asymptotic form of the vacuum polarizations $\Pi^I_\ell(s)$ for large $s$ is obtained from perturbative QCD together with IR information given by the vacuum condensates. The Finite Energy Sum Rules relate that behavior to integrals of the spectral density in the intermediate energy region. In that region the spectral density is saturated by two pion states and therefore agrees with the modulus squared of the form factor. By analyticity, the modulus of the form factor determines its phase. Since we also expect unitarity to be saturated, by Watson's theorem, the phase of the form-factor and partial waves are the same. In that way, the information from pQCD enters the bootstrap. 

\subsection{QCD parameters}

 Although it would be interesting to consider different theories and possibly compare with lattice calculations, we first want to compare pion bootstrap with experiment. In order to do that we use quark masses and data from \cite{narison_2004} based on experimental results. The information we need is \cite{narison_2004}:\footnote{In that reference, the values are: $m_u=3.6\,\MeV$, $m_d=6.5\,\MeV$  at $\epsilon=2\,\GeV$, and $\alpha_2=0.37$ at $\epsilon=m_c$.}
\beq\label{qcddata1}
 s_0=(1.2\,\GeV)^2,\ \ \alpha_s=0.4, \ \ \ m_u = 4\,\MeV \ \ \ m_d = 7.3\,\MeV
\eeq
and also the estimates
\beq\label{qcddata2}
 \langle \frac{\alpha_s}{\pi}G^2\rangle \simeq 0.023\,\GeV^4, \ \  \langle j_S(0) \rangle=m_q\langle \bar{u}u+\bar{d}d\rangle \simeq -(0.1\,\GeV)^4
\eeq
The parameters are evaluated at the renormalization scale $\mu=\sqrt{s_0}=1.2\,\GeV$. 
For light quarks the contribution from the condensates is small and probably only required for higher precision calculations in future work. We include it here for completeness \footnote{It might be also possible to use the bootstrap to put bounds on the values of the condensates which we leave for future work.}. From low energy QCD we need to use that $m_\pi\simeq 140\MeV$. It is convenient to set $m_\pi=1$ so all the other quantities are written in such units. We also use the value $f_\pi\simeq 92\,\MeV$ which is not an independent parameter since it could be computed\footnote{It is actually more precise to set the value $f_\pi\simeq 92\,\MeV$ and compute from there the quark condensate. In our procedure we only use $f_\pi$ to identify the region of parameters where we expect to find QCD.} from the previous data using the GMOR\cite{PhysRev.175.2195} relation $m_\pi^2 = -\frac{\langle j_S(0)\rangle}{f_\pi^2}$.      
Now we can write the numerical bounds for $s_0=1.2\,\GeV$:
\begin{equation}\label{srnumbers}
	\begin{aligned}
	\frac{1}{s_0^{n+2}} \int_4^{s_0} \rho^0_0(x) x^n dx &\simeq  3.09\times 10^{-8} \left\{\frac{27.38}{n+2}+0.61\ \delta_n +\ldots\right\} \\
	\frac{1}{s_0^{n+2}} \int_4^{s_0} \rho^1_1(x) x^n dx &\simeq -4.34\times 10^{-6} \left\{-\frac{13.26}{n+2}+0.41\ \delta_n +\ldots\right\} 
	\end{aligned} 
\end{equation}
or
\begin{subequations}\label{ffnumbers}
	\beqa
	\frac{1}{s_0^2} \intx|\cF^0_0|^2 \lesssim 4.42\times 10^{-7},\;&& \frac{1}{s_0^3} \intx x |\cF^0_0|^2 \lesssim 2.82\times 10^{-7}\;\;\;\; \\
	\frac{1}{s_0} \intx \frac{1}{x} |\cF^1_1|^2 \lesssim 5.76\times 10^{-5},\;&& \frac{1}{s_0^2} \intx |\cF^1_1|^2 \lesssim 2.70\times 10^{-5}\;\;\;\;
	\eeqa
\end{subequations}
These bounds are here for reference, in the bootstrap we bound the spectral density $\rho^I_\ell(s)$ with \eqref{srnumbers} and impose \eqref{Bmt}.  
Again, notice that everything is dimensionless since we set $m_\pi=1$ and that the contribution from the condensates is small. 

\section{Numerical implementation}

As described above, the gauge theory bootstrap involves the relatively new but by now standard S-matrix/form factor bootstrap, plus additional inputs from the IR -- properties resulting from chiral symmetry breaking, and UV -- properties of the asymptotically free theory through the SVZ sum rule. In this section, we describe the numerical implementation of each step.

\subsection{S-matrix bootstrap}

The numerical implementation of the S-matrix bootstrap follows a well-established procedure, namely to parameterize the scattering amplitudes incorporating the analyticity and crossing manifestly, and impose unitarity for the partial waves.  We briefly summarize the setup.

To numerically parameterize the scattering amplitude \eqref{Adef}, we map complex $\nu=s,t,u$ cut planes with cuts $\nu>4$ into a unit disk:
\begin{equation}\label{zmap}
	z(\nu)=\frac{2-\sqrt{4-\nu}}{2+\sqrt{4-\nu}}
\end{equation}
where the point $\nu=0$ is mapped to the center of the disk and the region $\nu>4$ above the cut is mapped to the upper half circle:
\begin{equation}\label{h33}
	z(\nu+i\epsilon)=e^{i\phi},\;\; \phi\in(0,\pi)
\end{equation}
with
\begin{equation}\label{h34}
	\nu(\phi)=\frac{8}{1+\cos(\phi)}
\end{equation}
We then discretize by taking $M$ points on the upper half circle
\begin{equation}\label{phii}
	\phi_i=\frac{\pi}{M}\bigg(i-\frac{1}{2}\bigg),\;\; i=1,2,\ldots,M
\end{equation}
so the parameters \eqref{Apara} takes the discrete set of variables:
\begin{equation}\label{Svars}
	\{T_0,\sigma_{\alpha,i},\rho_{\alpha,ij}\},\;\;\alpha=1,2,\; i,j=1,2,\ldots,M
\end{equation}
where $\rho_{2,ij}$ is symmetric in $i\leftrightarrow j$. The analytic partial waves can then be evaluated using the expression \eqref{fdef}. This is done both in the physical region $s>4$ for imposing the unitarity constraints \eqref{uni}, and in the unphysical region $0<s<4$ (where they are real) for imposing chiral symmetry breaking constraints as described in the following subsection. 

The parameterization used already satisfies crossing and analyticity, but the unitarity constraints reduce the space of allowed variables \eqref{Svars}. To get an idea of its properties we choose a few parameters that preferably depend on all the variables and plot their allowed space of values. In this case we choose simply two parameters $(f^0_0(3),f^1_1(3))$, the values of the partial waves continued to the unphysical region evaluated, as mentioned, from \eqref{fdef}. We then obtain a shape that contains all the allowed values as in fig.\ref{Sshape} below. Since the constraints define a convex space such shape is convex and can be mapped out by maximizing linear functionals. For example we can introduce a variable $t$, the extra constraints $f^0_0(3)=a+t \cos\alpha$, $f^1_1(3)=b+t \sin\alpha$ for some fixed $\alpha$ and a point $(a,b)$ inside the shape. If we maximize $t$, the maximum value of $t$ defines a point at the boundary of the allowed space. Sweeping the values of $\alpha\in[0,2\pi]$ gives the shape.

The second step of the method is to impose constraints from chiral symmetry breaking as we now do.

\subsection{Chiral symmetry breaking}  

 We assume chiral symmetry breaking and use the linearized form of the low energy amplitude \eqref{chiralratio}. This requires knowing the pion mass $m_\pi$ (here set to 1) and the pion decay constant $f_\pi$. To implement the requirement of chiral symmetry breaking, we consider the ratios of the $S0, P1, S2$ partial waves \eqref{chiralratio} in the unphysical low energy region $0<s<4$:
\begin{equation}\label{chiratio}
R^{\chi}_{21}(s)=\frac{f^2_0(s)}{f^1_1(s)}=\frac{3(2-s)}{s-4},\;\;R^{\chi}_{01}(s)=\frac{f^0_0(s)}{f^1_1(s)}=\frac{3(2s-1)}{s-4}
\end{equation}
These ratios are independent of $f_\pi$ and thus valid (within the linear approximation) for theories with different $f_{\pi}$ with the same symmetry breaking pattern. However, as commented in section \ref{chibreak}, when $f_{\pi}$ is small, the Goldstone boson is strongly coupled and the linear form of the partial waves at low energy fails. Therefore, requiring the ratios of the partial waves to match \eqref{chiratio} constrains the space of amplitudes to theories of Goldstone bosons interacting not too strongly.

In the unphysical region of $0<s<4$, we choose a few points $s_j$ and impose that 
\begin{equation}\label{chiralconstraints} 
	\begin{aligned}
		||f^0_0(s_j)-R^{\chi}_{01}(s_j)f^1_1(s_j)||&\le \epsilon^{\chi},\\
		||f^2_0(s_j)-R^{\chi}_{21}(s_j)f^1_1(s_j)||&\le \epsilon^{\chi},\;\; s_j\in(0,4)
	\end{aligned}
\end{equation}
with some norm and tolerance $\epsilon^\chi$. Setting different values of the tolerance we can find shapes in the plane $\big(f_0^0(s=3)$, $f_1^1(s=3)\big)$ that contain all allowed theories that satisfy the chiral symmetry constraints \eqref{chiralconstraints} up to the tolerance given. In our numerics, we choose the points to be $s_j=1/2,1,3/2,2$ and obtain shapes of the type shown in fig. \ref{chiral}. If the tolerance is too strict, only the free theory $f_0^0(3)=f_1^1(3)=0$ will be allowed. Therefore we pick a reasonable value that allows at least a region around the point given by \eqref{chiralratio} with the known value of $f_\pi$.  

Notice also that we impose the matching for values $s\le 2$ away from threshold and only the values of the ratios at four points. This means that the functions are not necessarily linear and therefore we are also not imposing the particular values of the scattering lengths predicted by chiral perturbation theory although its order of magnitude should be the same as the values of the amplitude at $s=3$ that we use to parameterize the shape. At this point we are also not using $f_\pi$ in the plot.

The next step is to use the form factor bootstrap for a selected number of partial waves, in this case $S0$ and $P1$.

\subsection{Form factor bootstrap and SVZ sum rules}

The form factor\footnote{We suppressed the isospin label $I$ in the rest of this paper since numerically we are using only two form factors $\mathcal{F}^{I=0}_0$ and $\mathcal{F}^{I=1}_1$.} $F_{\ell}(s)$ defined in \eqref{F0def} and \eqref{F1def} is an analytic function in the cut $s$-plane with cut $s>4$. It
can therefore be parameterized in terms of its imaginary part on the real axis $x>4$ as in \eqref{Fdisp}.
 The function in \eqref{Fdisp} can be mapped to a function on the unit disk through the map \eqref{zmap} above and parametrized by $\text{Im}F$ at discrete points on the upper half circle:
\begin{equation}\label{h35}
\text{Im}F_{\ell,i}=\text{Im}F_{\ell}(\nu(\phi_i)),\;\; i=1,2,\ldots M
\end{equation}
whereas the real part is computed as
\begin{equation}\label{h36}
\text{Re}F_{\ell,i}=1+K_{ij}\text{Im}F_{\ell,j},\;\; i,j=1,\ldots,M
\end{equation}
with \cite{He:2018uxa}:
\begin{equation}\label{h37}
	\begin{aligned}
	K_{ij} &=\tilde{K}_{i+j-2M-1}-\tilde{K}_{i-j},\;\; i,j=1,\ldots,M\\
	\tilde{K}_{m}&=\begin{cases}
		0,&m=0\\
		\frac{1-(-1)^m}{2M}\cotan\left(\frac{m\pi}{2M}\right),&\text{otherwise}\\
	\end{cases}
	\end{aligned}
\end{equation}
We can then compute the values $\mathcal{F}_{\ell,i}$ defined in \eqref{curF} where the factors between $\mathcal{F}_{\ell,i}$ and $F_{\ell,i}$ are evaluated at the points $s_i,\;\; i=1,\ldots,M$.

Finally, we have the spectral density $\rho_\ell(s)$ evaluated at the points
\begin{equation}\label{h39}
\rho_{\ell,i}=\rho_{\ell}(s(\phi_i))
\end{equation}
as variables.

To summarize, the S-matrix/form factor bootstrap is numerically parameterized by the following discrete set of variables:
\begin{equation}\label{h40}
\{T_0,\sigma_{\alpha,i},\rho_{\alpha,ij}, \text{Im}F_{\ell,i},\rho_{\ell,i}\}
\end{equation}
with $\alpha=1,2,\; i,j=1,\ldots,M$, $\ell=0,1$.

In the simplest setup we consider in this paper, we impose the constraints of the positive semidefinite matrices:
\begin{equation}\label{positiveB}
\left(
\begin{array}{c c c}
1 &\ S_{\ell,i}\ &\ \ \cF_{\ell,i} \\
S^*_{\ell,i} &\ 1\ & \ \ \cF^*_{\ell,i}\\
\cF^*_{\ell,i} &\ \cF_{\ell,i}\ &\ \ \rho_{\ell,i}
\end{array}
\right) \succeq 0
\end{equation}
for the $S0$ and $P1$ waves, whose UV information on the form factor and the spectral density is extracted from perturbative QCD (see below); for other partial waves, we simply impose
\begin{equation}\label{h41}
\left(
\begin{array}{c c c}
1 &\ S_{\ell,i}\\
S^*_{\ell,i} &\ 1
\end{array}
\right)\succeq 0
\end{equation}
but of course, in a more involved computation, one can use the constraint \eqref{positiveB} with more UV information on the other $I,\ell$.
\medskip

The strong interaction energy scale is input in terms of the value of the coupling $\alpha_s(s_0)\simeq0.4$ evaluated at $s_0=1.2\,\GeV$. We then compute the finite energy sum rule \eqref{srnumbers} by discretizing the integral as
\beqa\label{srnum}
\int_4^{s_0}dx \rho(x)x^n \rightarrow \frac{\pi}{M}\sum_{i=1}^M\bigg(\frac{ds}{d\phi}\bigg)_i s_i^n\rho_{i}
\eeqa  
where
\begin{equation}\label{h42}
\frac{ds}{d\phi}=\frac{8\sin \phi}{(1+\cos\phi)^2}
\end{equation}
and impose the sum rules \eqref{srnumbers} as linear constraints on the bootstrap. We use $n=0,1$ for the $S0$ wave and $n=-1,0$ for the $P1$ wave.
In practice, we impose the sum rule by allowing a tolerance
\begin{equation}\label{h43}
||\frac{\pi}{M}\sum_{i}\big(ds/d\phi\big)_i s_i^n\rho_{i}-\text{QCD value}||\le \epsilon^{\text{SR}}
\end{equation}
to guarantee the feasibility of the numerical problem. The QCD values are those from  \eqref{srnumbers} and the tolerance is chosen such that enough information of the sum rule is put into the bootstrap yet not too strictly to make the problem infeasible.

\medskip

Finally, the form factor at high energy has the asymptotic behavior \eqref{ffQCD} although in our numerical implementation, we only need that it goes zero at large $s$. As argued in eq. \eqref{ffratioasymp}, the form factors of scalar and vector currents have the same expected asymptotic behavior up to a constant depending on the charge. We therefore impose the condition on the rescaled form factor \eqref{curF}:
\beq\label{FFasym}
||\mathcal{F}_0(s_i)||^2 \lesssim 2 m_q^2\, \epsilon^{FF}, \ \ \ \ \ ||\mathcal{F}_1(s_i)||^2\lesssim \half \epsilon^{FF}, \ \ s_i>s_0
\eeq
for some $\epsilon^{FF}$ whose order of magnitude we can estimate as follows: The spectral density always provides an upper bound and the form factor is decreasing. Therefore we get:  
\begin{equation}\label{h44}
\rho_1(s=s_0)\sim \mathcal{O}(10^{-3}) \Rightarrow |F_{\pi}(s\ge s_0)|\lesssim \mathcal{O}(10^{0})
\end{equation}
where we included the appropriate numerical factor. On the other hand, we expect that the pQCD result \eqref{ffQCD} provides a lower bound since the form factor might not have decreased all the way to its asymptotic value when $s=s_0$. We then get an interval
\beq  \label{h43b}
 \mathcal{O}(10^{-2}) \lesssim|F_{\pi}(s\ge s_0)|\lesssim \mathcal{O}(10^{0})
\eeq
within which we can make a reasonable initial choice of $\epsilon^{FF}$ to use in \eqref{FFasym} for constraining the asymptotic behavior of the form factor. Numerically we can refine this choice, as usual, by making sure that the constraints reduce the allowed space without making the problem unfeasible.    
Finally, due to the factor between $F$ and $\mathcal{F}$ as in \eqref{curF} (which we evaluate at $s=s_0$), the $\epsilon^{FF}$ is taken to be
\begin{equation}
\epsilon^{FF}\simeq \mathcal{O}(10^{-5})
\end{equation}

\section{Results}

Now we describe the numerical results as they apply to QCD. We emphasize that the numerical inputs for this calculation are $N_c=3$, $N_f=2$, $\alpha_s(s_0=1.2\,\GeV)=0.4$, the pion mass $m_\pi=140\,\MeV$, the pion decay constant $f_{\pi}\simeq 92\,\MeV$, the quark masses in \eqref{qcddata1} and the quark and gluon condensates in \eqref{qcddata2}. The condensates make only a very small contribution to the results but we include them for completeness. The only parameter that we could change is $s_0$, or, equivalently the value of $\alpha_s$ at which we match high and low energy. Adjusting it, can lead to better agreement with experiment, specially on the $\rho$ mass. However we prefer to remain agnostic and not use the data we want to match. As mentioned, we are not trying to do QCD phenomenology but to test a general bootstrap procedure for finding the low energy physics of gauge theories.      

\subsection{S-matrix bootstrap}

\begin{figure}
	\centering
	\includegraphics[width=0.8\textwidth]{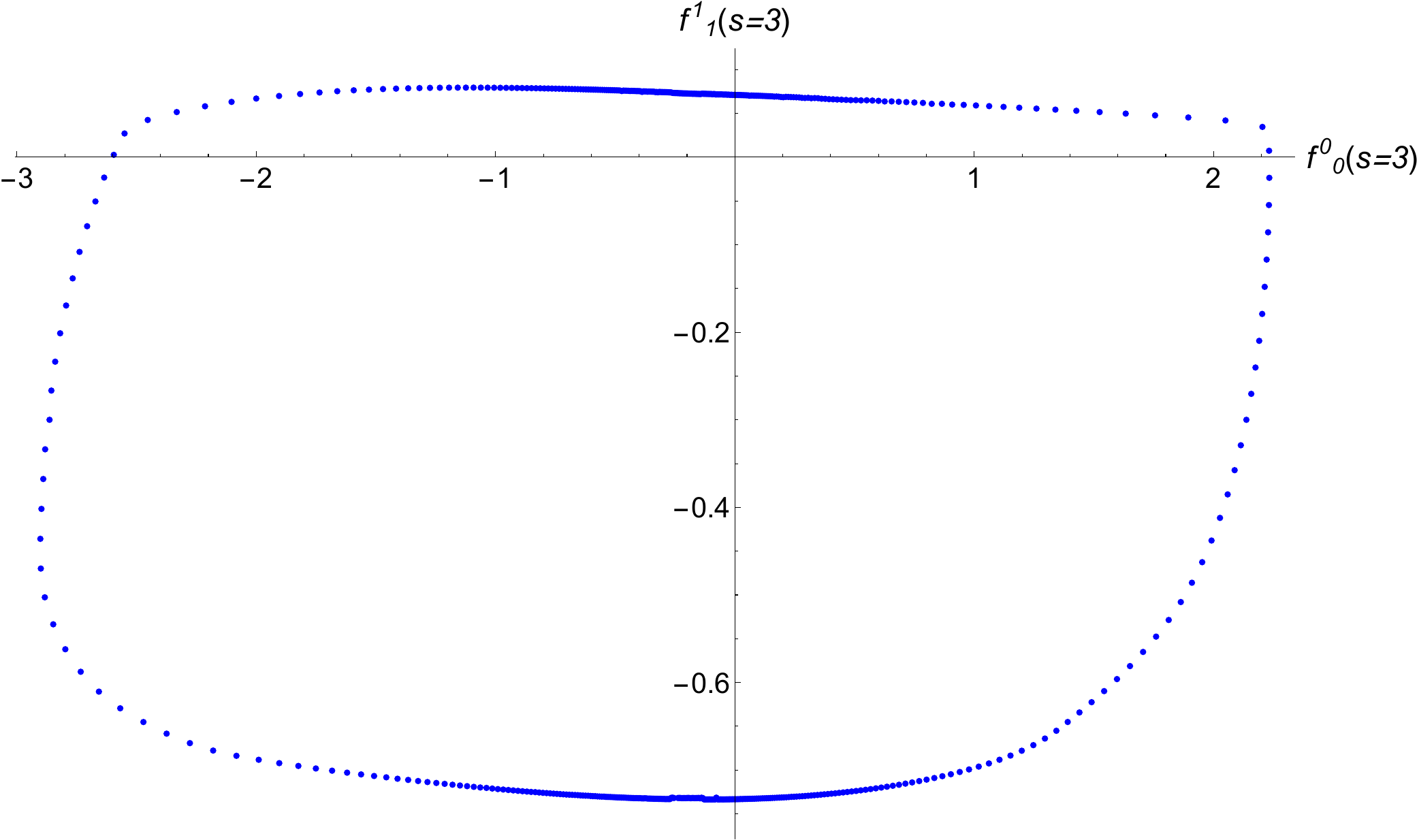}
	\caption{The space of amplitudes projected onto a plane parameterized by $f^0_0(s=3)$ and $f^1_1(s=3)$ as a result of pure S-matrix bootstrap with the constraints of analyticity, crossing and unitarity.}
	\label{Sshape}
\end{figure}

 In the initial step we want to characterize the space of S-matrices allowed by analyticity, crossing, unitarity, and the global $SU(2)_V$ symmetry. As discussed above, to visualize such space, we consider a two-dimensional projection of the infinite dimensional space of amplitudes to a plane parameterized by $f^0_0(s=3)$ and $f^1_1(s=3)$. This choice of parameters is due to our setup that focuses on the $S0$ and $P1$ partial waves where we input explicit UV information from perturbative QCD.
The resulting shape is depicted in figure \ref{Sshape}, where we have used $M=50$ in \eqref{phii} for discretization and imposed unitarity for 10 partial waves per isospin. Inside the shape are all the possible values that $f^0_0(s=3)$ and $f^1_1(s=3)$ can have under the constraints of analyticity, crossing and unitarity. 
It is perhaps important to indicate that the boundary points are associated with specific amplitudes. For the interior points there is an infinite number of amplitudes that attain those values at $s=3$. Such space of amplitudes is certainly interesting to study in itself, but we will move on to focus on the specific theory of pions, the shape in fig. \ref{Sshape} will play no further role.

\subsection{Chiral symmetry breaking}

\begin{figure}
	\centering
	\includegraphics[width=0.9\textwidth]{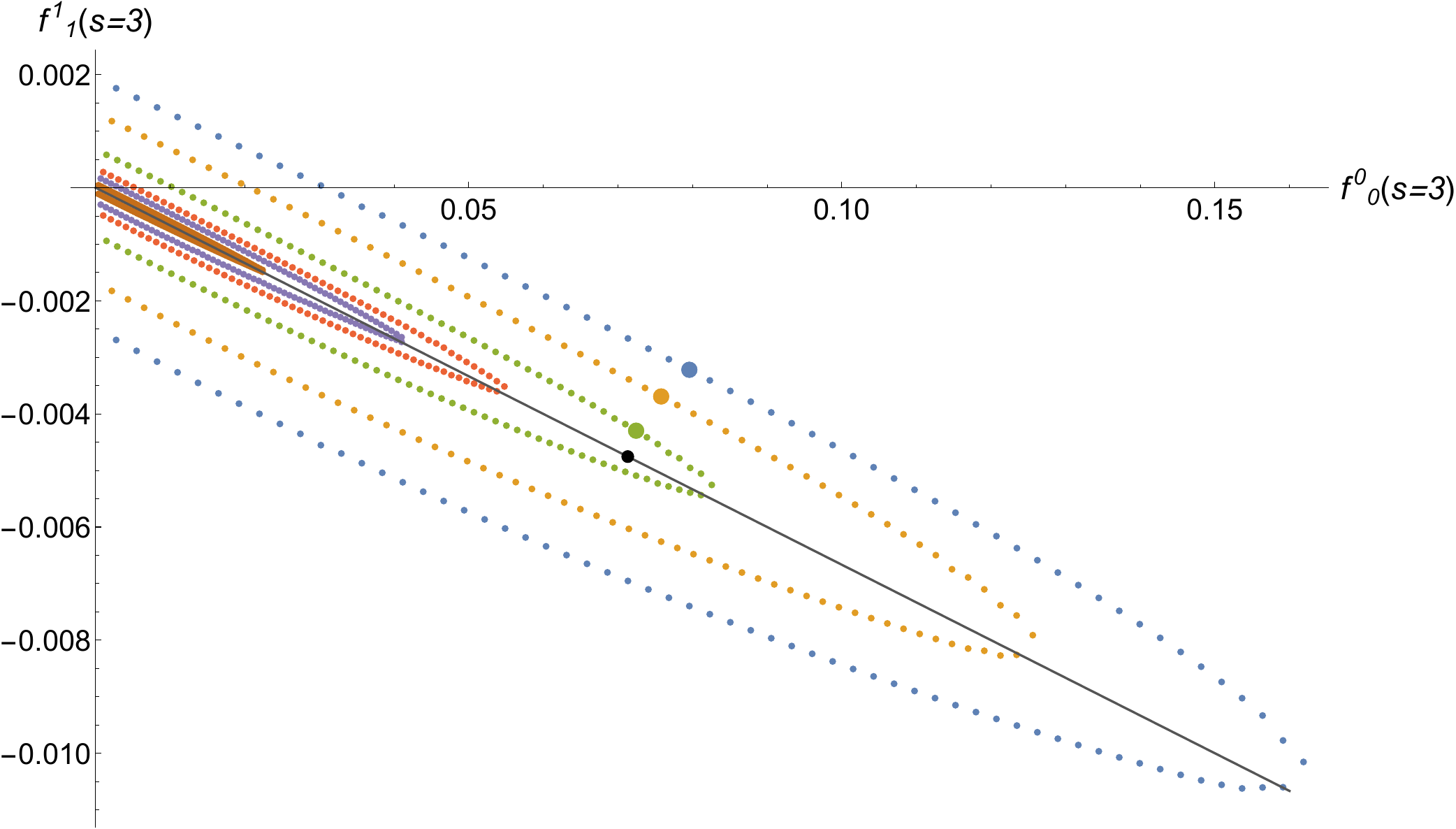}
	\caption{We plot the allowed space (only the relevant $f_0^0(3)>0$ side) restricted by the chiral constraints \eqref{chiralconstraints} with tolerances $\epsilon^{\chi}=6\times10^{-3},4\times10^{-3},2\times10^{-3},1\times10^{-3},6\times10^{-4},2\times10^{-4}$ (from the outer shape inward). The black line are the values given by the linear Weinberg model \eqref{h5} with varying values of $f_\pi$ and the black dot the one with $f_\pi\simeq\, 92\MeV$. We want to impose the constraints without excluding that point. A few highlighted points are chosen to explore the partial waves in the unphysical region as plotted in fig \ref{linearity}.}
	\label{chiral}
\end{figure}

\begin{figure}[h]
	\centering
	\begin{subfigure}{0.48\textwidth}
		\raggedright
		\includegraphics[width=\textwidth]{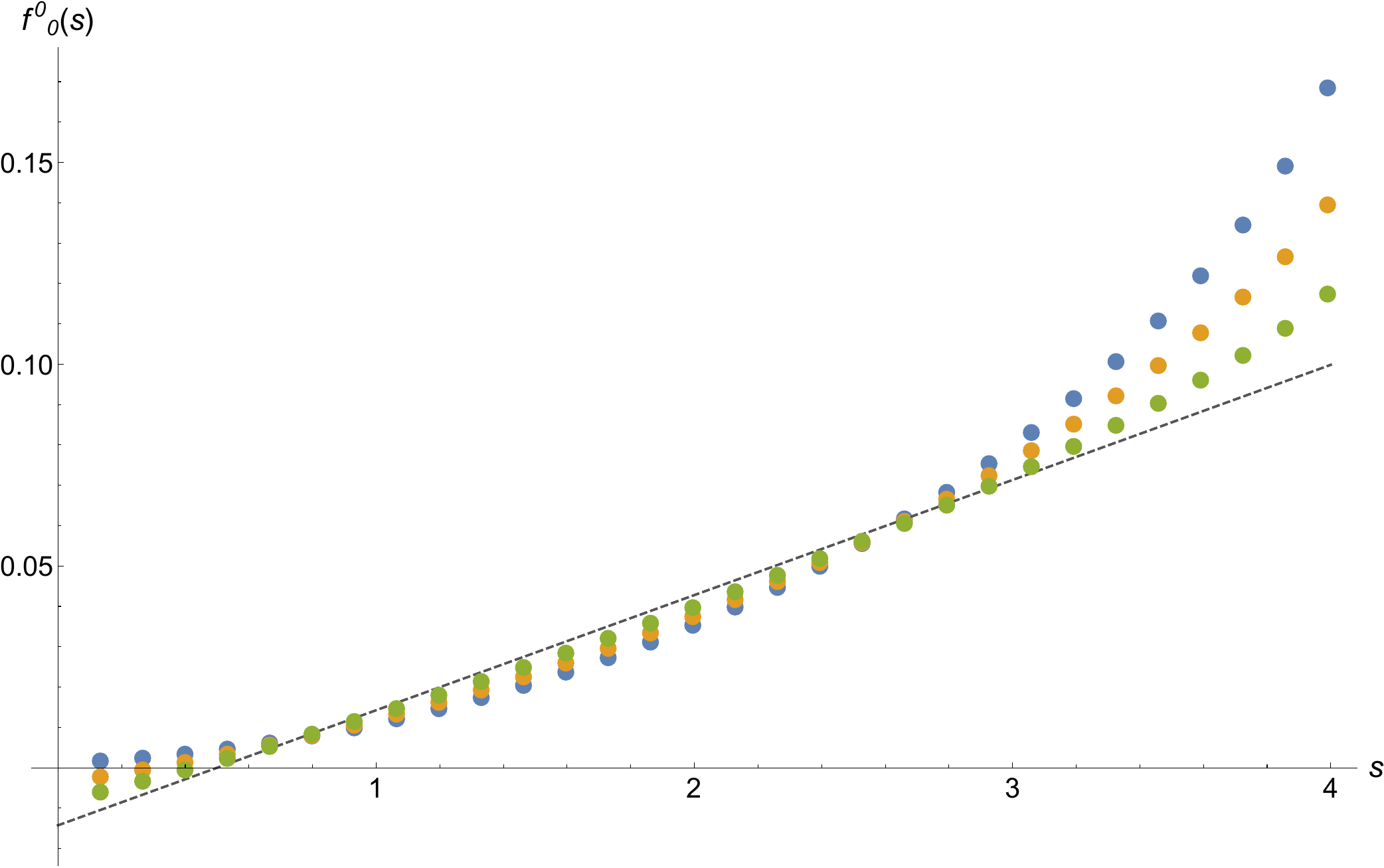}
		\caption{S0}
	\end{subfigure}
	\begin{subfigure}{0.48\textwidth}
		\centering
		\includegraphics[width=\textwidth]{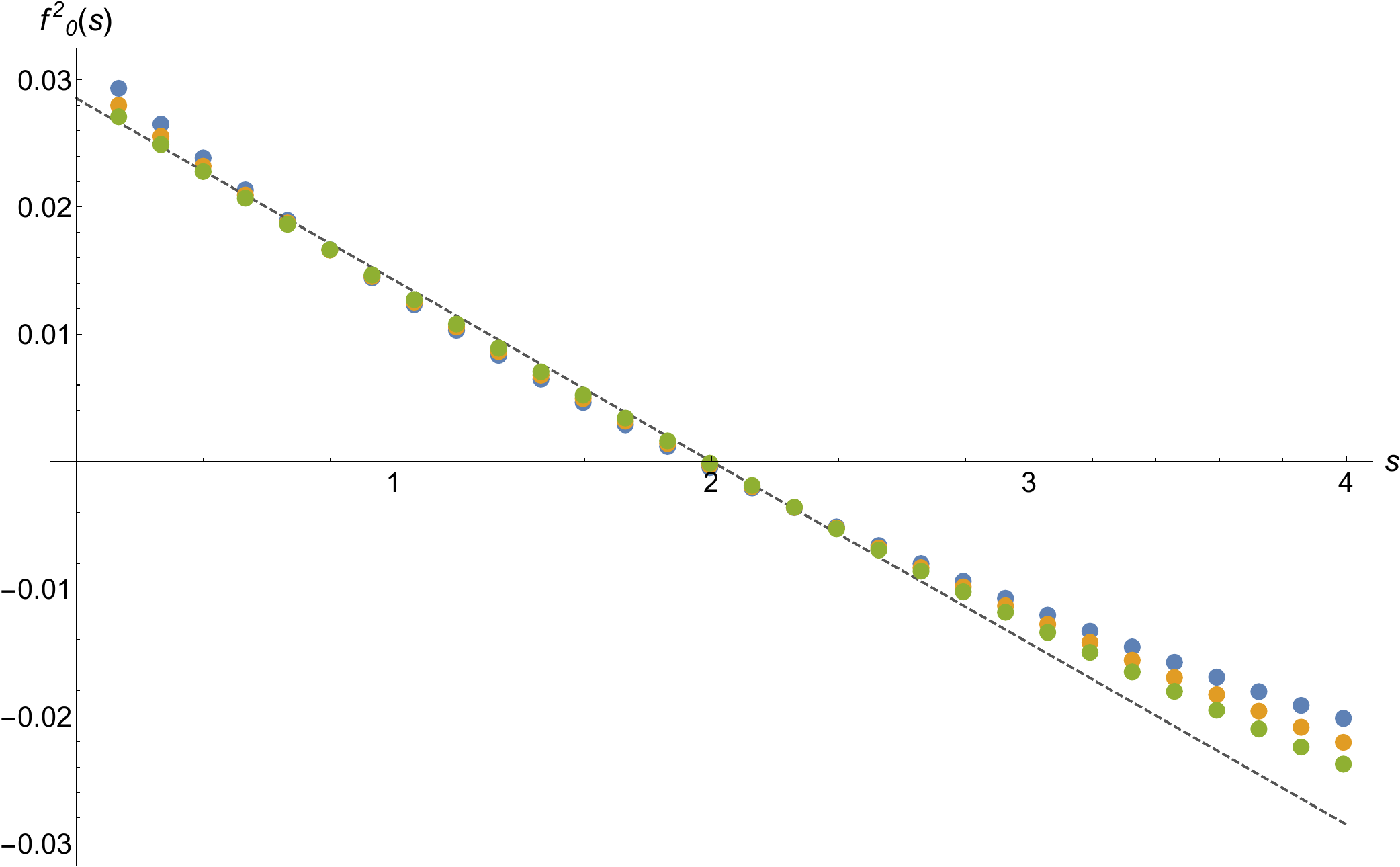}
		\caption{S2}
	\end{subfigure}\\
		\begin{subfigure}{0.48\textwidth}
		\centering
		\includegraphics[width=\textwidth]{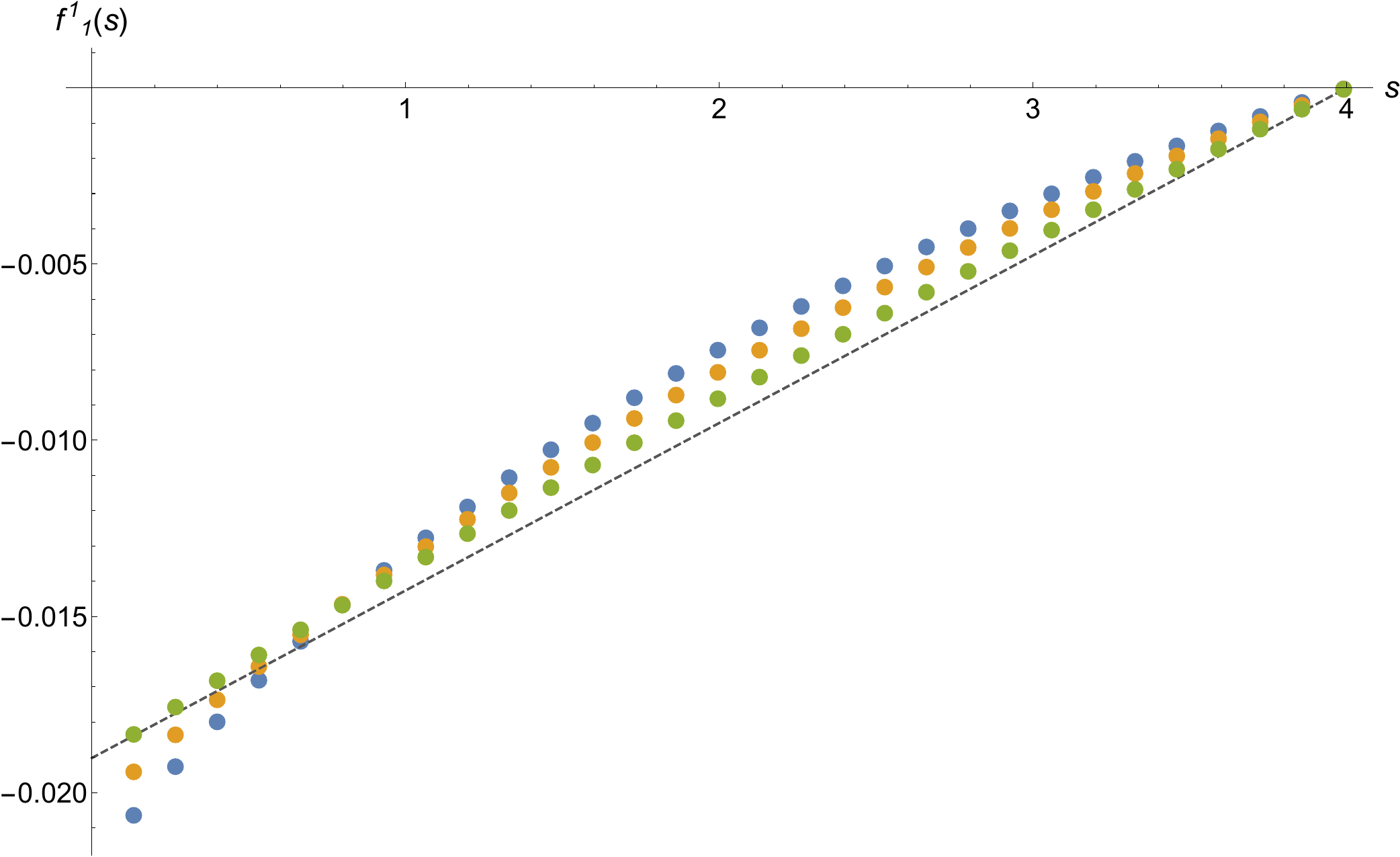}
		\caption{P1}
	\end{subfigure}
	\caption{Partial waves in the unphysical region $0<s<4$ for the points highlighted in fig.\ref{chiral}. We compare them with the linear approximation (dashed line) and find that the green points match better the linear prediction without excluding the physical value of $f_\pi$ in fig.\ref{chiral}.}
	\label{linearity}
\end{figure}

The way we impose the chiral symmetry breaking constraints is through requiring that the chiral ratios \eqref{chiratio} at the four points $s=1/2,1,3/2,2$ agree with the numerical values within a certain tolerance, as in \eqref{chiralconstraints}. Restricting the values of these ratios does not imply that the amplitudes are linear. However,  even though $s=3$ is not one of the points chosen to fix the ratios, the values of $f^0_0(s=3)$ and $f^1_1(s=3)$ become very restricted and the allowed shape is reduced significantly as seen in fig. \ref{chiral} where we used different tolerances to understand how the space is reduced (only the $f_0^0(3)>0$ side where we expect to find pion scattering is shown).
In fact, the space of amplitudes shrinks to a thin region around the linear order chiral prediction (arbitrary $f_\pi$):
\begin{equation}\label{h45}
f_1^1(s=3)=-\frac{1}{15}f_0^0(s=3)
\end{equation}
depicted in the figure as a black line. The black dot in the figure indicates the prediction of \eqref{chiralratio} for the physical value of $f_\pi=92$MeV. 
For the larger tolerances the partial waves are not approximately linear in $0<s<4$ and therefore we discard them. This is shown in fig.\ref{linearity} where the partial waves were plotted in the unphysical region $0<s<4$ with colors corresponding to those in fig.\ref{chiral}. Notice that in fig.\ref{chiral} we highlighted three points whose partial waves are plotted in fig.\ref{linearity}. The most notable deviation from linearity is in the $S0$ wave since the chiral zero of the amplitude disappears for the blue points. There is certain ambiguity in how much deviation from linearity is acceptable so we choose among the smallest deviation that does not exclude the physical value of $f_\pi$ (black dot). The value $\epsilon^\chi=0.002$ (green points), that we now choose,  allows the physical value of $f_\pi$ while at the same time make the partial waves match better the constraints as seen in fig.\ref{linearity}.  We do not expect small changes in this tolerance to affect the results. The main thing is that we have a region of allowed values in the vicinity of the black dot.    

\begin{figure}
	\centering
	\includegraphics[width=0.9\textwidth]{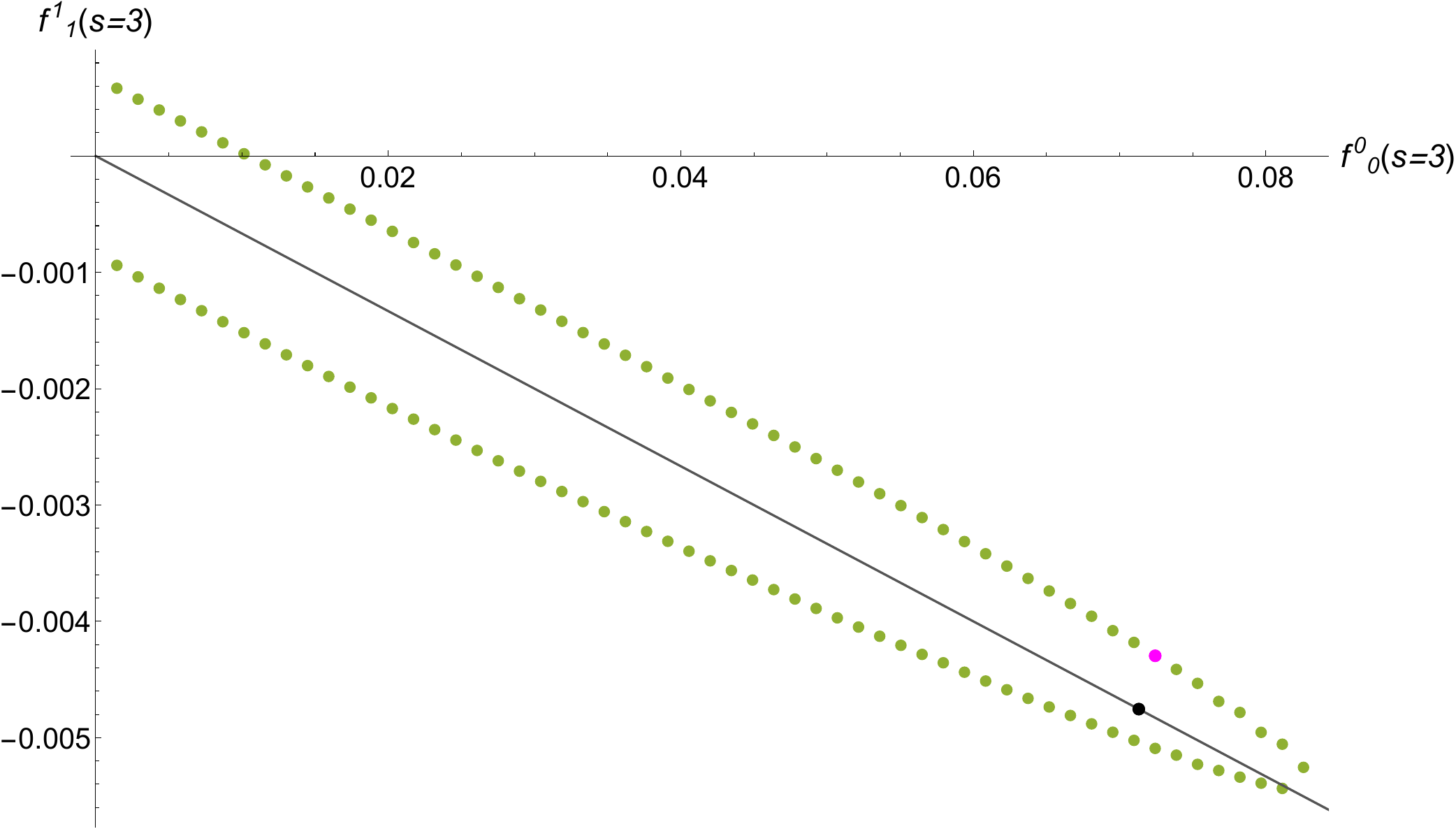}
	\caption{Following fig. \ref{chiral} we select tolerance $\epsilon^\chi=2\times10^{-3}$. We choose a point closest to the black dot to explore the partial waves in the physical region. Notice that only points at the boundary (green points) have partial waves associated with them. The partial waves at the magenta point and other nearby agree very well with experimental values for the $S0$ and $S2$ waves but not for the $P1$ as can be seen in fig. \ref{chiS0S2phaseshift}. }
	\label{chiral2}
\end{figure}
\begin{figure}[h]
	\centering
	\begin{subfigure}{0.48\textwidth}
		\raggedleft
		\includegraphics[width=\textwidth]{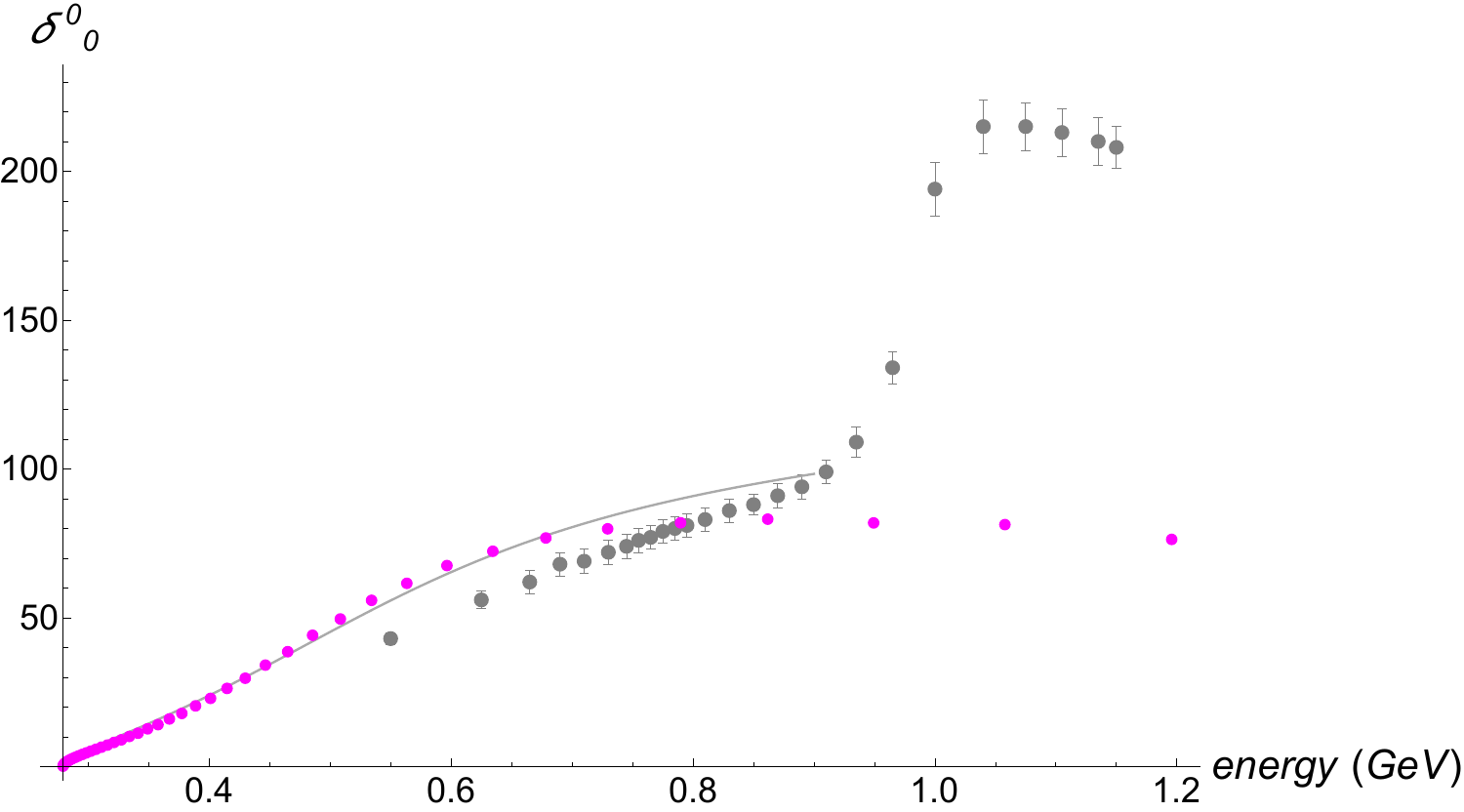}
		\caption{S0}
	\end{subfigure}
	\begin{subfigure}{0.48\textwidth}
		\raggedright
		\includegraphics[width=\textwidth]{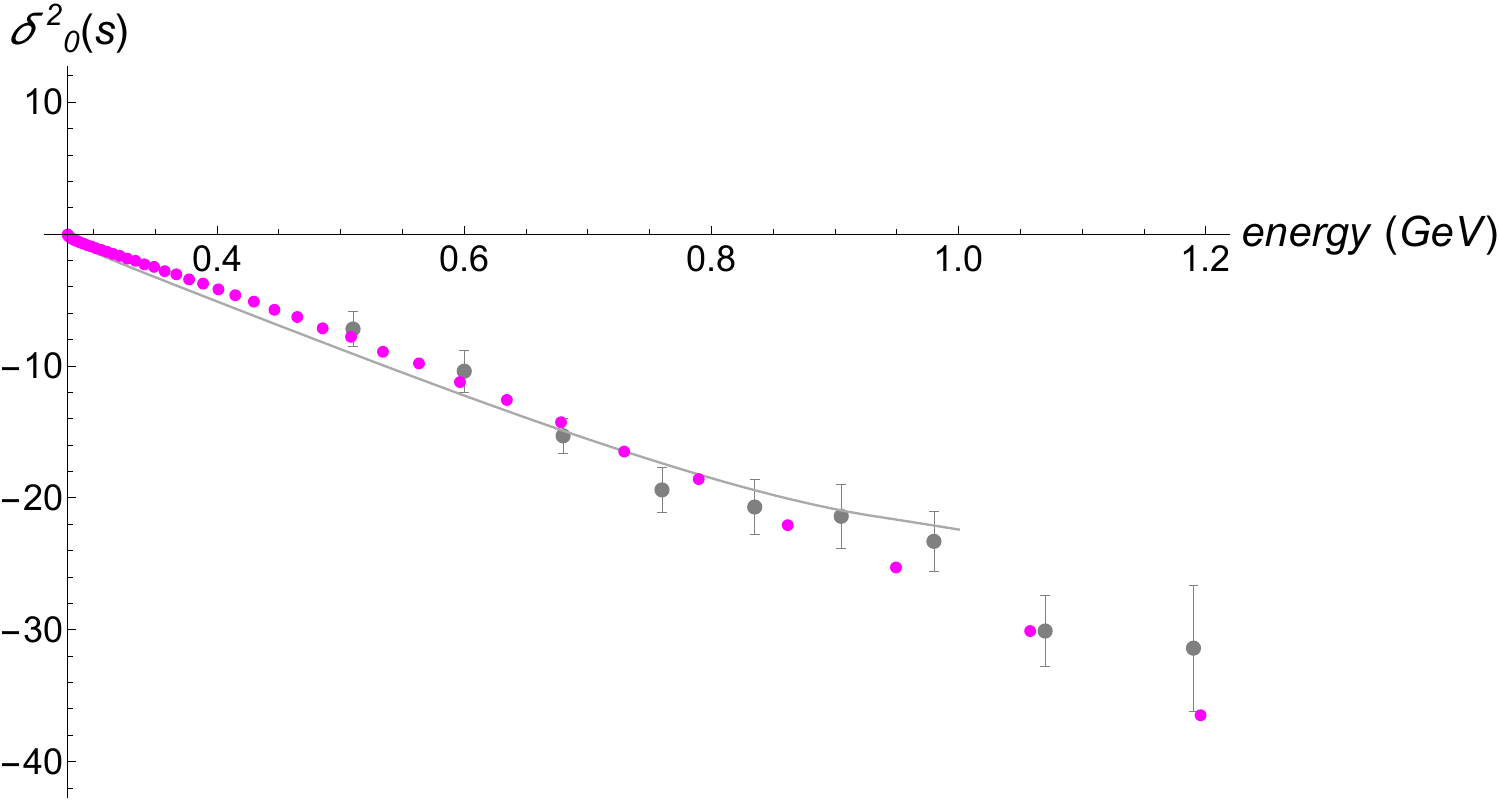}
		\caption{S2}
	\end{subfigure}\\
	\begin{subfigure}{0.48\textwidth}
		\centering
		\includegraphics[width=\textwidth]{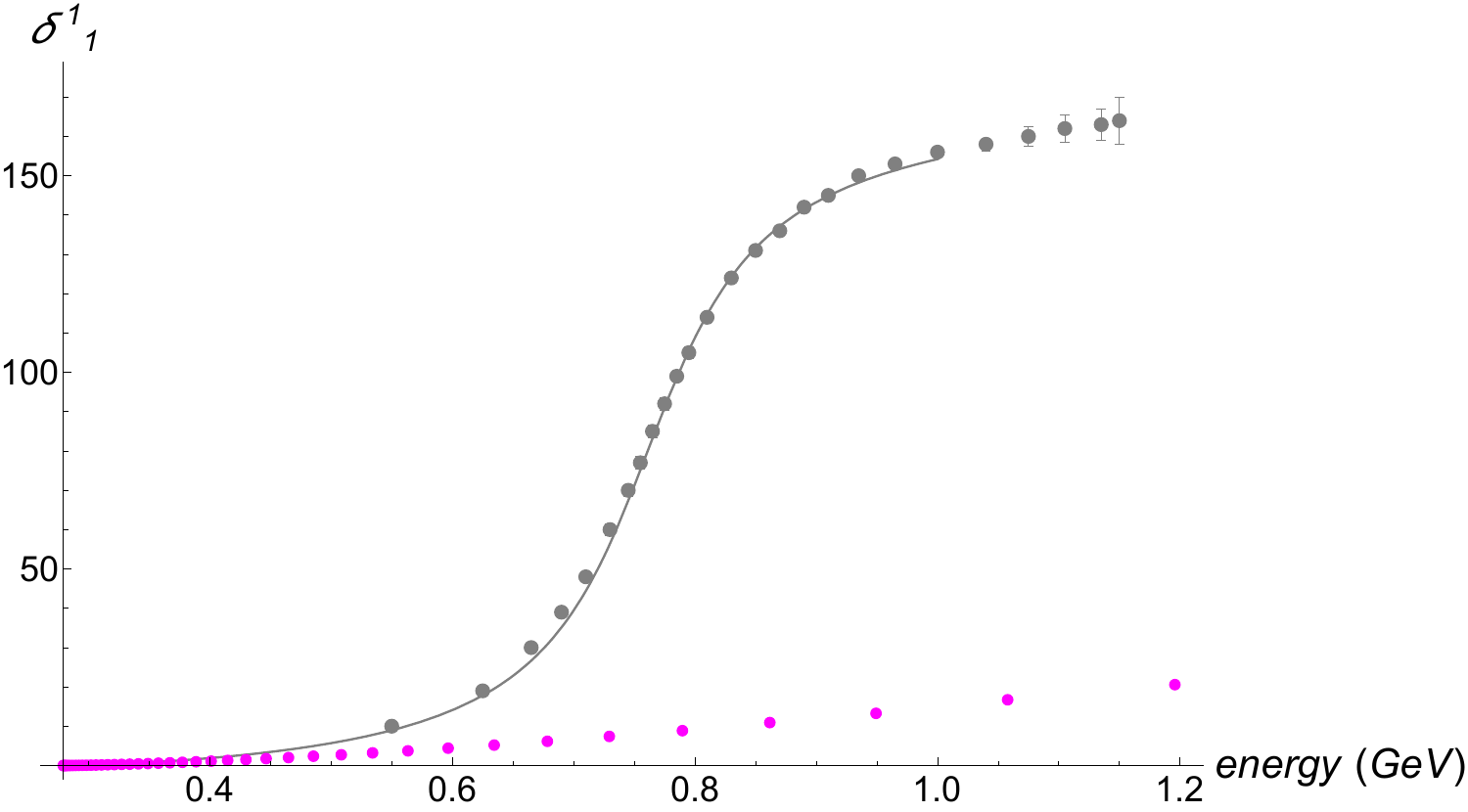}
		\caption{P1}
	\end{subfigure}
	\caption{Phase shifts of the $S0,S2$ and $P1$ waves in the physical region of the experimental results \cite{Protopopescu:1973sh,LOSTY1974185} (gray dots), a phenomenological fit \cite{Pelaez:2004vs} (gray dashed line) and the bootstrap with (only) chiral constraints (magenta points). The $S0$ and $S2$ waves are correct but the $P1$ evidently not, it requires further input from QCD. Energy is in $\GeV$ for ease of comparison with other work.}
	\label{chiS0S2phaseshift}
\end{figure}

 We remark that the scales of fig.\ref{chiral} and fig.\ref{Sshape} are quite different, it is clear that the weakly coupled pion scattering amplitudes occupy a very tiny region near the origin -- the free theory. Note that moving along the thin shape in fig. \ref{chiral} corresponds to choosing different values of $f_{\pi}$. The value of $f_\pi$ decreases as we get far from the origin. Since the shape terminates, we see that the (approximate) linear form of the low energy partial waves imply a lower bound on the range of possible $f_{\pi}$ as we have argued earlier. Namely a low $f_\pi\ll m_\pi$ implies that even at low energy the pions are strongly coupled and the linear approximation would not be valid. Notice also that once again, at the boundary of the shape we have well defined partial waves, namely the partial waves that attain the given values at $s=3$ while at the same time satisfying our imposed chiral constraints with the tolerance given. In particular we can choose a point (magenta color, see fig. \ref{chiral2}) near the chiral point (black) and plot the amplitudes as we show in fig. \ref{chiS0S2phaseshift}. They are remarkably compatible with experimental results for the $S0$ and $S2$ waves but not the $P1$ wave. Similar results are found for other nearby points. This suggests that, to obtain $P1$ we need more information that can only come from the UV in the form of the sum rules. It also makes the prediction that the $S0$ and $S2$ waves are the same for other gauge theories within a range of number of colors, quark masses and coupling constants as long as the low energy is reasonably well approximated by chiral perturbation theory and the $S0$ sum rule does not modify this conclusion. For example, if we increase $N_c$ by a large amount the sum rules affect the $S0$ wave more and a narrow resonance might develop. The idea that the overall shape of the $S0$ and $S2$ waves is determined largely by chiral perturbation theory is in fact expected already from the Weinberg model but the bootstrap gives more evidence and also the actual shape.  
 
This ends our discussion of the chiral constraints. For future work, it would be interesting to improve this setup by using chiral perturbation theory to systematically incorporate small deviations from the linear form.  
\bigskip
\subsection{SVZ sum rules}

\begin{figure}[h]
	\centering
	\includegraphics[width=0.9\textwidth]{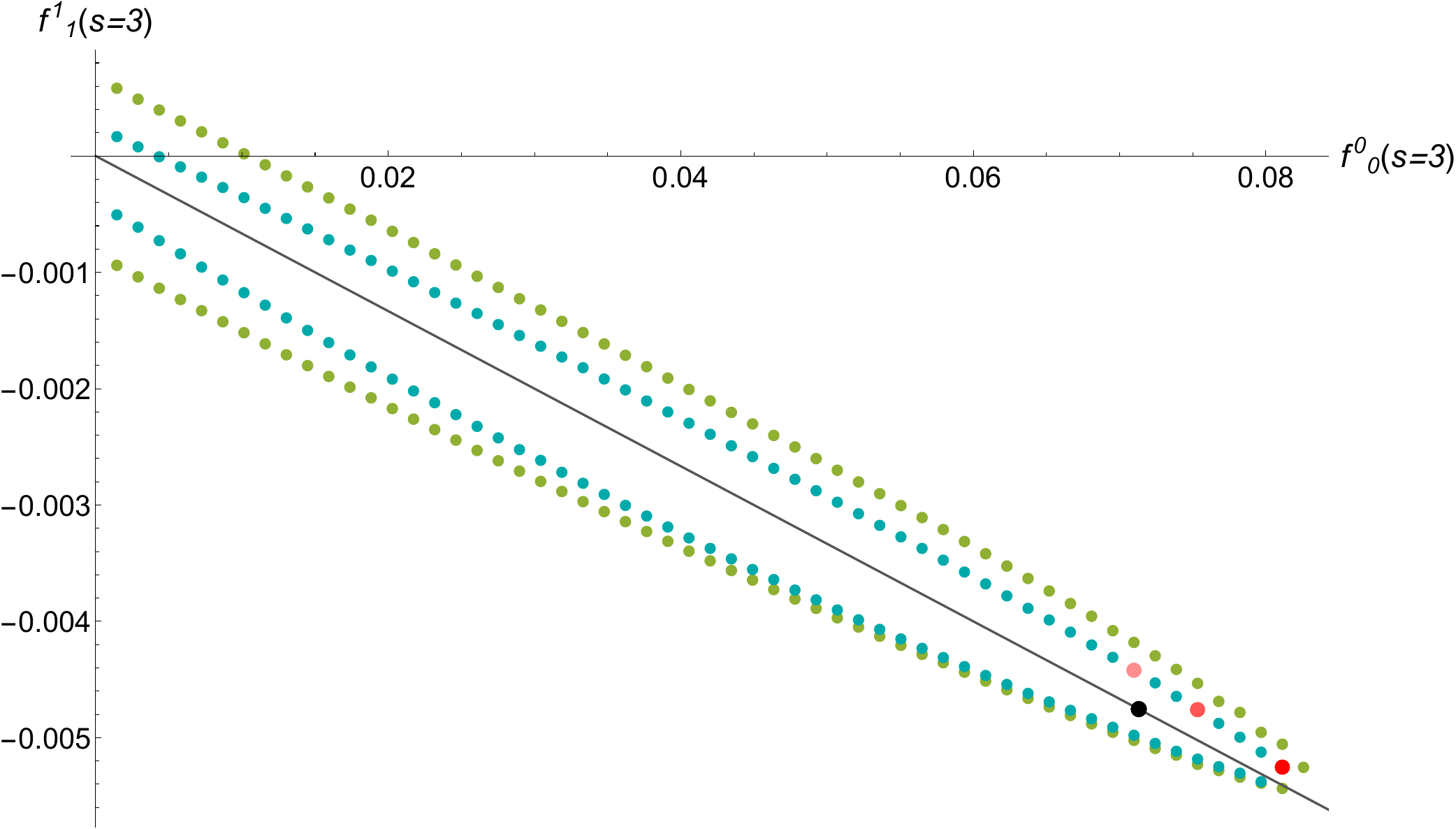}
	\caption{The shape after adding the SVZ finite energy sum rules. The upper boundary shrinks notable but the lower not so much. We expect that the upper boundary has acquired QCD information. }
	\label{srfig}
\end{figure}

\begin{figure}
	\centering
	\includegraphics[width=0.8\textwidth]{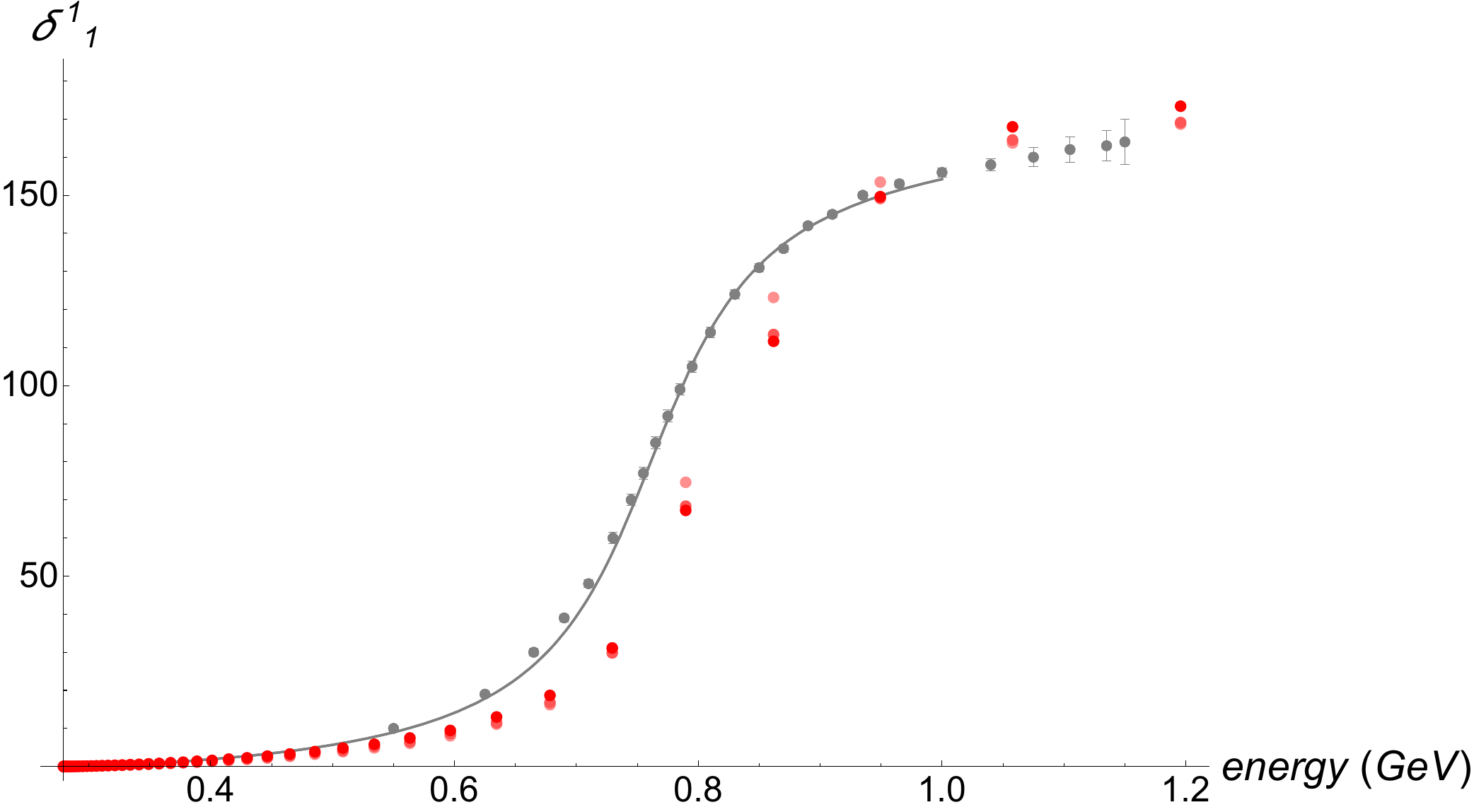}
	\caption{Bootstrap results for the $P1$ wave phase shift after using the QCD sum rules are indicated with red dots. Experimental data (gray dots) on the phase shift is taken from \cite{Protopopescu:1973sh,LOSTY1974185}. Gray dashed line indicate the phenomenological fit to the experiment from \cite{Pelaez:2004vs}. The resonance is shifted by a few percent but the shape is quite good. Notice that we did not use any parameters to fit the experiment other than the known pion and quark masses together with $\lqcd$ and $f_\pi$. Energy is in $\GeV$ for ease of comparison with other work.}
	\label{P1phaseshift}
\end{figure}

\begin{figure}[h]
	\centering
	\begin{subfigure}{0.54\textwidth}
		\raggedleft
		\includegraphics[width=1\linewidth]{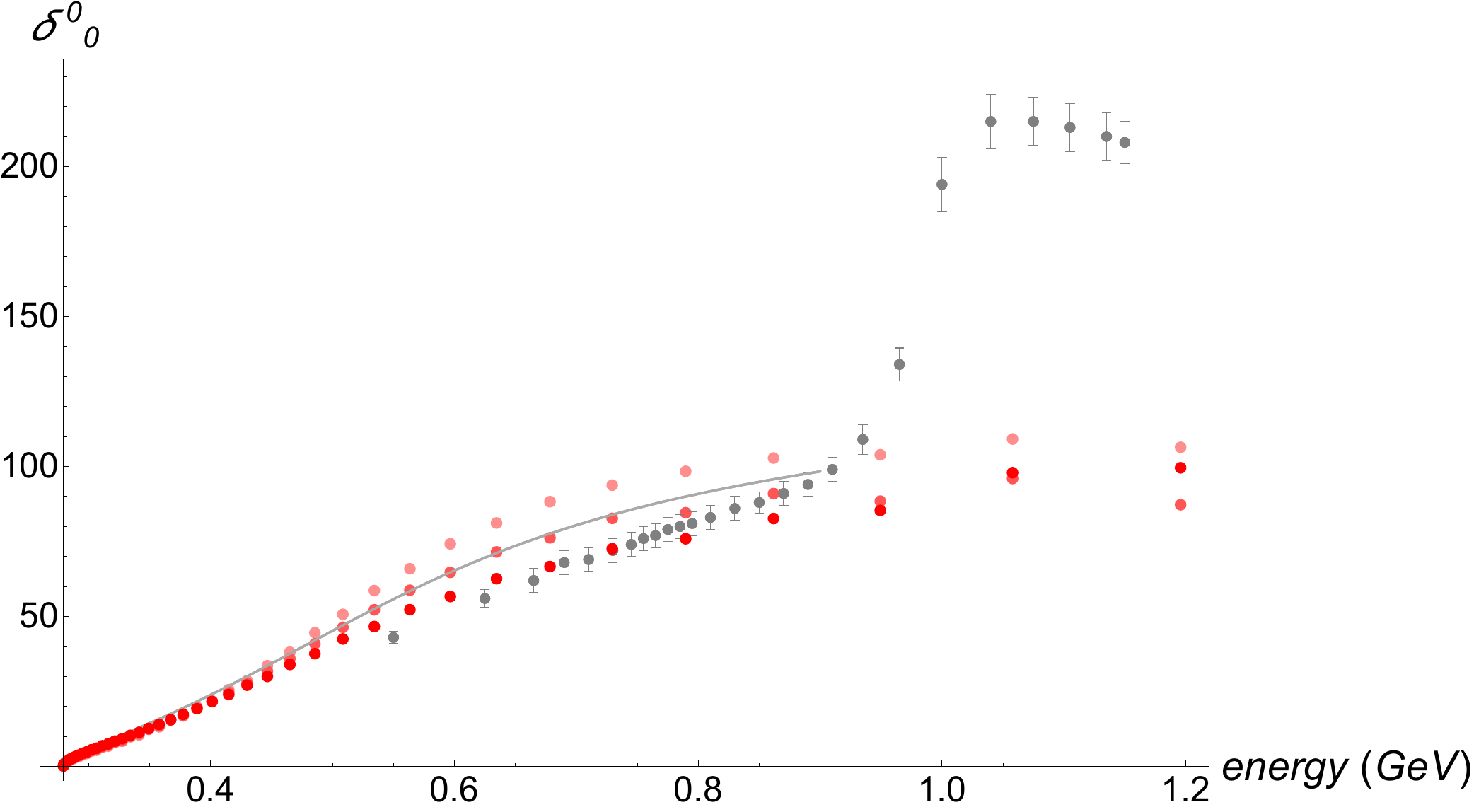}
		\caption{S0}
	\end{subfigure}
	\begin{subfigure}{0.54\textwidth}
		\raggedright
		\includegraphics[width=1\linewidth]{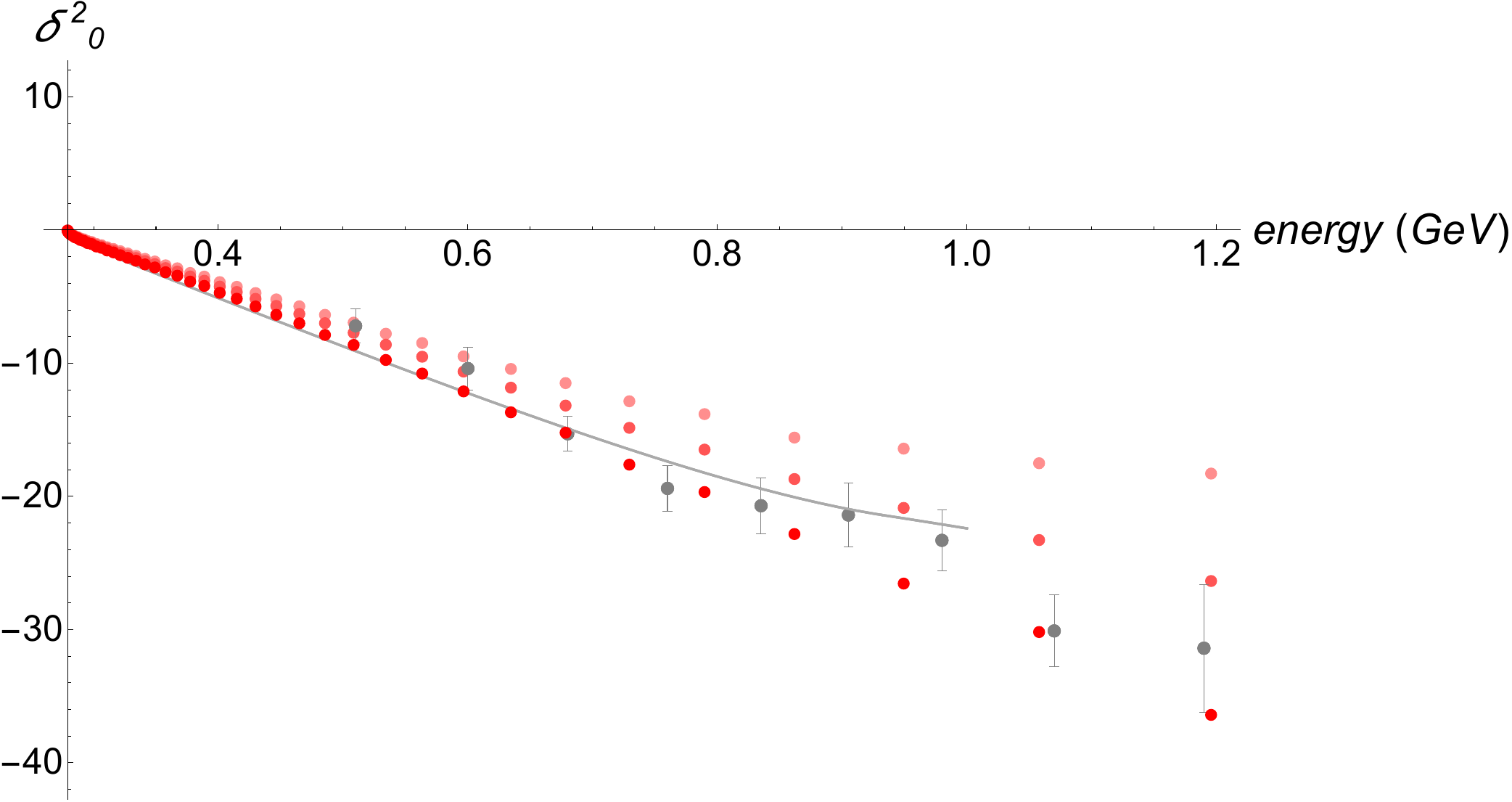}
		\caption{S2}
	\end{subfigure} 
		\caption{Bootstrap results are indicated with red, pink, light pink dots. Experimental data on the phase shift is taken from \cite{Protopopescu:1973sh}, \cite{LOSTY1974185}. Gray dashed line indicate the phenomenological fit to the experiment from \cite{Pelaez:2004vs}. Choosing different points in fig. \ref{srfig} we obtain similar but not equal phase shifts. Energy is in $\GeV$ for ease of comparison with other work.}
		\label{S0S2phaseshift}
\end{figure}

Finally, in the last step we impose the sum rules \eqref{srnumbers} (two per partial wave, $n=0,1$ for $S0$ wave and $n=-1,0$ for $P1$ wave) and the asymptotic behavior of the form factors \eqref{FFasym}. This step is where actual information from the UV theory, namely perturbative QCD is incorporated. They constrain the asymptotic behavior of the form factor when combined with the positivity condition \eqref{Bmt}. The result -- using the same tolerance for the chiral part -- is depicted in figure \ref{srfig}. More specifically, the plots are produced by taking $\epsilon^{SR}=2\times 10^{-3}$ in \eqref{h43} and $\epsilon^{FF}=6\times 10^{-5}$.
The shape (cyan points) shrinks a bit with respect to the previous shape of fig. \ref{chiral} (green points). This makes sense. While the low energy constraints significantly reduce the shape, there are still infinitely many possible scattering amplitudes contained within this thin shape, which differ from each other in terms of higher energy behaviors and this cannot be simply visualized in such a 2d projection. The sum rule and asymptotic behaviors of the form factor however helps navigate the bootstrap to the particular theories in these alternative dimensions of parameters. Upon imposing such UV constraints, we land on the low energy amplitudes corresponding to the correct UV theories. Notice that the upper curve is modified much more than the lower one. This means that the sum rules constrain more the upper points and we expect to get better results there. This, however is a discrete choice between upper and lower boundary, we have no particular way of choosing between them. Previous experience with the bootstrap would suggest looking at the at the tip of the shape (here the red dot). However the results seem quite robust so we choose two other points (pink, light pink) near the chiral point (black dot).       
 The phase shifts for those points depicted in fig.\ref{P1phaseshift} and fig.\ref{S0S2phaseshift} agree reasonably well with experiment including the $\rho$ meson resonance in the $P1$ channel.  The resonance energy where the phase shift crosses $\pi/2$ is slightly shifted from the real world data on the mass of the rho at $770\,\MeV$ by roughly $6\%$. This might be improved by choosing other value of $s_0$. However choosing parameters to fit the experiment seems against the spirit of the paper where we want to use a minimum of input. On the other hand the shape of the resonance seems quite good. In fact, given that we are considering the $N_f=2$ flavor quark models and considering the lowest terms in the perturbative QCD vacuum polarization, we consider the results to be compatible with the real world QCD. It would of course be interesting to take into account more data from perturbative QCD in future work. 

Notice that the three bootstrap (red, pink, light pink) curves in fig.\ref{S0S2phaseshift} coincide at low energy but they spread at high energy. This indeed suggest that they are determined from the low energy side.  For the $P1$ wave instead in fig.\ref{P1phaseshift} they coincide both at low and high energy but hey spread in the middle indicating that they are determined by low and high energy data. 

 In summary, the phase shifts are quite reasonable, in fact we optimistically believe they are \emph{very good}, the agreement with experiment is remarkable given the little information we used and the fact that QCD is obviously not just an $N_c=3$, $N_f=2$ gauge theory.  One way to go from here would be to incorporate the strange quark and isospin symmetry breaking to find $K\bar{K}$ production in the $S0$-wave. In any case, for the moment we believe it will also be useful to understand other gauge theories with different values of $N_c$ and $N_f$ but still with exact flavor symmetry.  

To summarize, the FESR and the asymptotic suppression of the form factor is clearly important to constrain the S-matrix/form factor bootstrap to find the signature of QCD in the low energy -- the rho resonance. Notice that this has been argued phenomenologically, for example see an illuminating discussion \cite{Colangelo:2001df}. There it is argued that the shape of the $S0$ wave is a consequence of pion low energy dynamics whereas the $\rho$ is a consequence of QCD. This agrees with our findings since the QCD sum rules were mostly useful to get the $\rho$.  

\section{Conclusions}

    In this paper we proposed a bootstrap method to solve the well known problem of finding the low energy physics of an asymptotically free gauge theory with massive quarks that undergoes confinement and chiral symmetry breaking. In such case the low energy theory is described by a scalar field theory of pions, and the main problem is to compute the pion scattering matrix. Applying the method to the important case of $N_c=3$ and $N_f=2$ we find partial waves ($S0$, $S2$ and $P1$) in good agreement with experimental data indicating that the method indeed works, at least to compute those QCD-pions partial waves.   In particular the $S0$ and $S2$ waves are largely independent of the high energy data, whereas the $P1$ is determined by the UV information. 
    This also means that, although the original idea of the bootstrap called for just imposing the general constraints of analyticity, crossing and unitarity, those constraints are clearly not enough to identify what S-matrix corresponding to a given UV theory, in particular the high energy data is already needed for the $P1$ wave. To incorporate information on the UV theory we started with the method recently proposed by Karateev, Kuhn and Penedones to introduce the form factors and the spectral density into the bootstrap. Then we incorporated the UV information using a finite energy version of the SVZ sum rules to relate perturbative QCD computation of the two current correlators to the spectral density at intermediate energies where it is saturated by two pion states. This information seems enough to determine the $P1$ wave.     
    
     Optimistically, the method will extend to more partial waves, find more resonances, and also help solve other models. In particular it will be interesting to investigate the dependence on $N_c$ and attempt other values of $N_f$ where results could be compared with lattice computations.
     It will also be interesting to include the gauge theory constraints in the case of massless pions studied in \cite{Guerrieri:2020bto}.   
      Another point is that, as mentioned in the text, at high energies the spectral density is no longer saturated by the two particle states. The difference between the spectral density and the form factor is related to particle production and could be a basis to include particle production in the bootstrap. In any case, it will be great to further explore and expand this framework to gain new understanding of gauge theories at low energy.
	
\section{Acknowledgements}

 M.K. wants to thank LPENS and Shanghai Tech for hospitality while this work was being done. We both want to thank Swissmap Les Diablerets Research Station for hospitality while this work was being completed. We are grateful to J. Penedones, B. Van Rees, P. Vieira and specially to A. Guerrieri  for various discussions on the S-matrix bootstrap and applications to pion physics. We would also like to thank L. Cordova, K. H\"aring, M. Paulos, S. Rychkov, A. Shapere, A. Zhiboedov for interesting comments on a previous version of this work. We would also like to thank the referee for comments that improved the paper. 
 
 This work was supported in part by DOE through grant DE-SC0007884 and the QuantiSED Fermilab consortium.

\appendix

\numberwithin{equation}{section}

\section{Test on numerical parameters}

In figs. \ref{P1phaseshift} and \ref{S0S2phaseshift}, we have shown the bootstrap results obtained for the $S0,P1,S2$ phase shifts in reasonable agreement with experimental data. The numerical calculation is done by discretizing the energy range using $M=50$ interpolation points and imposing unitarity on $L=10$ partial waves per isospin. This choice is due to a compromise between precision and speed. In this appendix, we briefly explore other values of $M$ and $L$ to understand how sensitive the results are to these numerical parameters.    

\begin{figure}[H]
	\centering
	\begin{subfigure}{0.7\textwidth}
		\centering
		\includegraphics[width=1\linewidth]{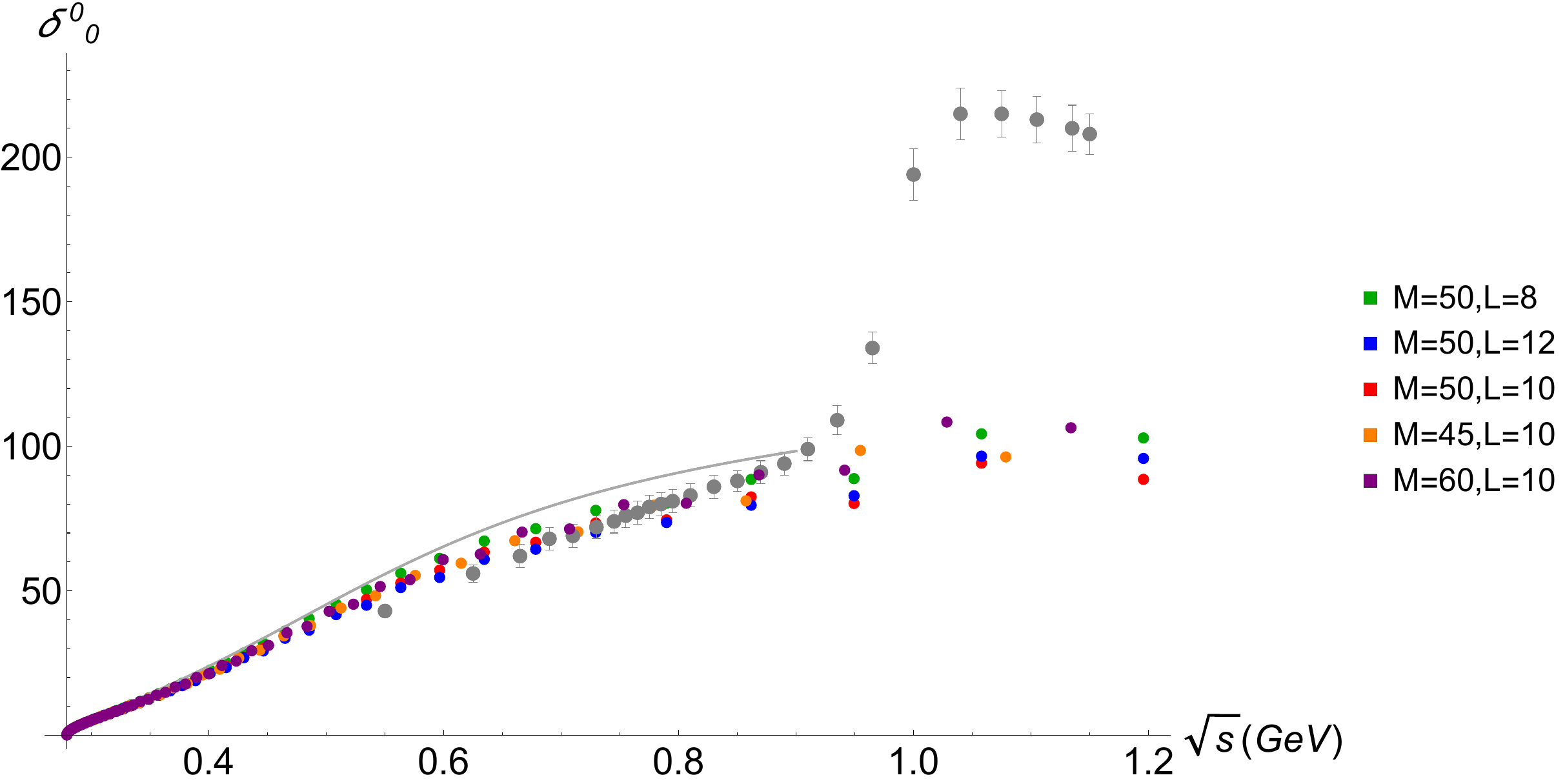}
		\caption{S0}
	\end{subfigure}
	\begin{subfigure}{0.7\textwidth}
		\centering
		\includegraphics[width=1\linewidth]{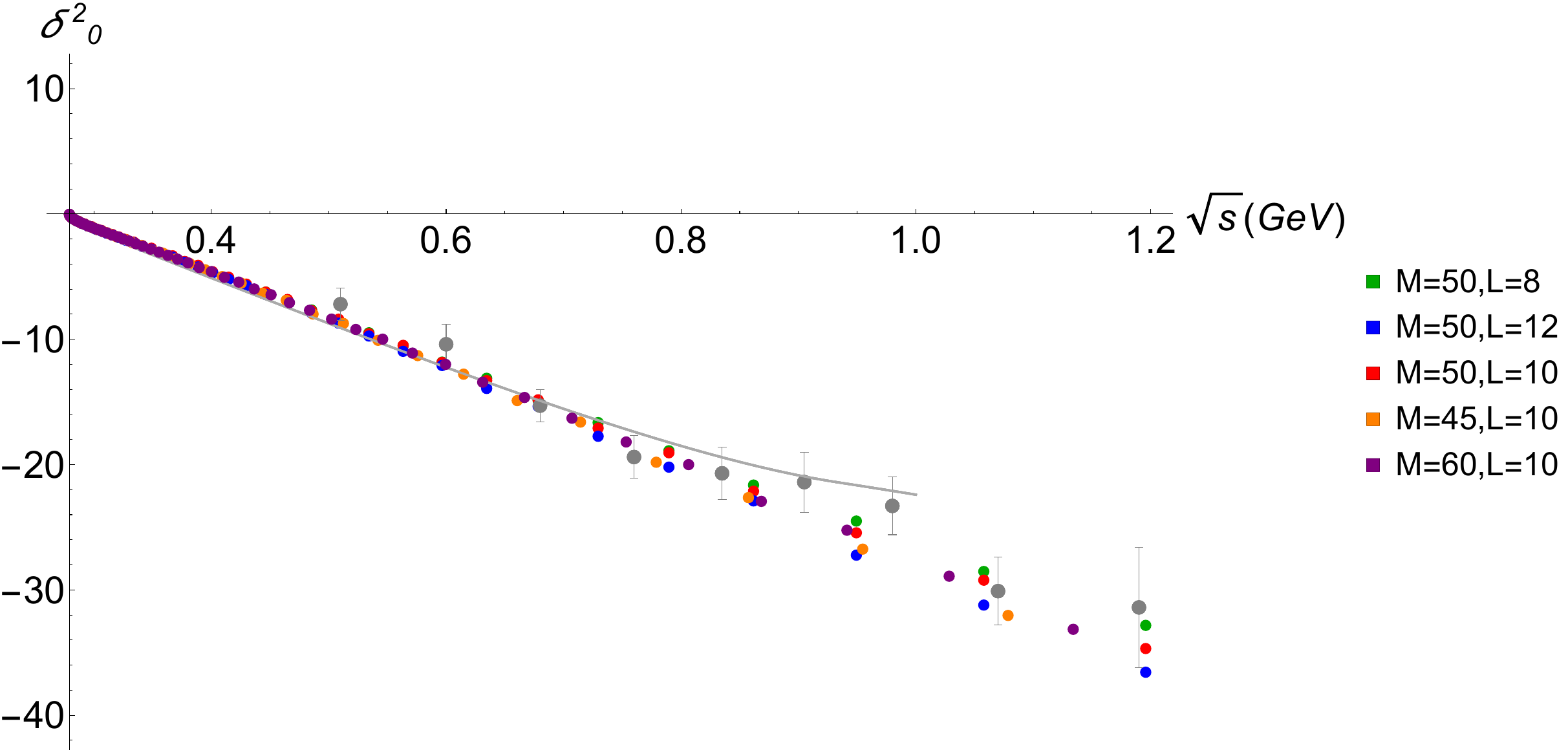}
		\caption{S2}
	\end{subfigure} 
	\begin{subfigure}{0.7\textwidth}
		\centering
		\includegraphics[width=1\linewidth]{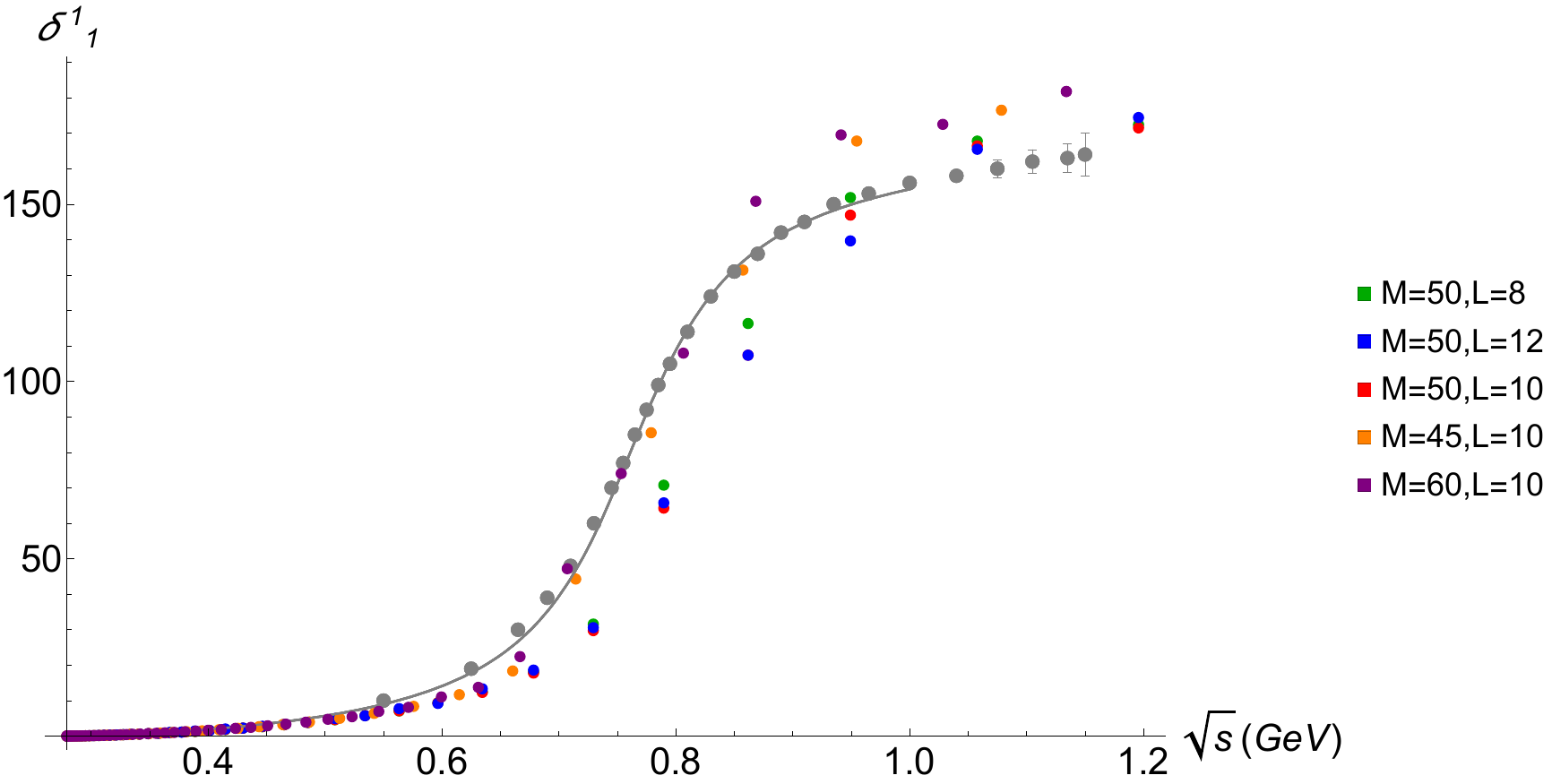}
		\caption{P1}
	\end{subfigure} 
	\caption{}
	\label{MLcomparison}
\end{figure}

In fig. \ref{MLcomparison}, we plot the $S0,P1,S2$ phase shifts for different values of $M$ and $L$. For $M=50$, we have plotted the results with $L=8$, $10$, $12$. In all three partial waves there is a reasonably good convergence. The situation with fixed $L=10$ and varying $M=45$, $50$, $60$ is slightly more complicated. Note that due to our numerical implementation, modifying $M$ changes both the number of bootstrap variables as well as the unitarity and SVZ sum rules constraints. We see from the plots that the $S0, S2$ phase shifts have a very good convergence, while the $P1$ phase shifts show a $\rho$ resonance with the peak slightly shifted depending on $M$.

\bibliographystyle{utphys}
\bibliography{references}

\end{document}